\documentclass[10pt, aps, prd, amsmath, floats, floatfix, twocolumn, notitlepage,
superscriptaddress, nofootinbib, showpacs, longbibliography]{revtex4-2}
\usepackage[T1]{fontenc}
\usepackage[utf8]{inputenc}
\usepackage{lmodern}
\usepackage{verbatim}
\usepackage{physics}
\usepackage{orcidlink}
\usepackage[all]{hypcap}
\usepackage{hyperref}
\hypersetup{
    colorlinks=true, 
    linkcolor=linkcolor, 
    citecolor=linkcolor, 
    filecolor=linkcolor, 
    urlcolor=linkcolor
}
\definecolor{linkcolor}{rgb}{0.0,0.0,1.0}
\usepackage{graphicx}
\usepackage{xspace}
\usepackage{amssymb}
\usepackage{amsmath}
\usepackage{bm}
\usepackage{microtype}
\usepackage{booktabs}
\usepackage[english]{babel}
\usepackage{blindtext}
\usepackage{array}
\usepackage{tabularx}
\usepackage{multirow}
\usepackage{subfigure}
\usepackage{fontawesome}
\usepackage{mathrsfs}
\usepackage{mathtools}
\usepackage{natbib}

\begin{document}

\title{Comprehensive analysis of time-domain overlapping \\ gravitational wave transients: A Lensing Study}

\author{Nishkal Rao~\orcidlink{0009-0006-4551-7312}}%
 \email{nishkal.rao@students.iiserpune.ac.in}
\affiliation{%
 Department of Physics, Indian Institute of Science Education and Research, Pashan, Pune - 411 008, India
}%
\affiliation{%
 Inter-University Centre for Astronomy and Astrophysics (IUCAA), Post Bag 4, Ganeshkhind, Pune 411 007, India
}%
\author{Anuj Mishra~\orcidlink{0000-0002-2580-2339}}%
 \email{anuj.mishra@icts.res.in}
 \affiliation{%
 International Centre for Theoretical Sciences, Tata Institute of Fundamental Research, Bangalore 560089, India
}%
\affiliation{%
 Inter-University Centre for Astronomy and Astrophysics (IUCAA), Post Bag 4, Ganeshkhind, Pune 411 007, India
}%
\author{Apratim Ganguly~\orcidlink{0000-0001-7394-0755}}%
 \email{apratim@iucaa.in}
\affiliation{%
 Inter-University Centre for Astronomy and Astrophysics (IUCAA), Post Bag 4, Ganeshkhind, Pune 411 007, India
}
\author{Anupreeta More~\orcidlink{0000-0001-7714-7076}}%
 \email{anupreeta@iucaa.in}
\affiliation{%
 Inter-University Centre for Astronomy and Astrophysics (IUCAA), Post Bag 4, Ganeshkhind, Pune 411 007, India
}%
\affiliation{%
 Kavli IPMU (WPI), UTIAS, The University of Tokyo, Kashiwa, Chiba 277-8583, Japan
}

\begin{abstract}
Next-generation gravitational-wave (GW) detectors will produce a high rate of temporally overlapping signals from unrelated compact binary coalescences. Such overlaps can bias parameter estimation (PE) and mimic signatures of other physical effects, such as gravitational lensing. In this work, we investigate how overlapping signals can be degenerate with gravitational lensing by focusing on two scenarios: Type-II strong lensing and microlensing by an isolated point-mass lens. We simulate quasicircular binary black-hole pairs with chirp-mass ratios $\mathscr{M}_{\rm B}/\mathscr{M}_{\rm A}\in\{0.5,\,1,\,2\}$, signal-to-noise ratio (SNR) ratios $\mathrm{SNR}_{\rm B}/\mathrm{SNR}_{\rm A}\in\{0.5,\,1\}$, and coalescence-time offsets $\Delta t_{\rm c}\in[-0.1,\,0.1]~\mathrm{s}$, and extend to a population analysis. Bayesian PE and fitting-factor studies show that the Type-II lensing hypothesis is favored over the unlensed quasicircular hypothesis ($\log_{10}\mathscr{B}^{\rm L}_{\rm U}>1$) only in a small region of the overlapping parameter space with $\mathscr{M}_{\rm B}/\mathscr{M}_{\rm A}\gtrsim1$ and $|\Delta t_{\rm c}|\leq0.03~\rm{s}$, with the inferred Morse index clustering near $n_j\simeq0.5$, indicative of Type-II lensing, for the cumulative study. Meanwhile, false evidence for microlensing signatures can arise because, to a reasonable approximation, the model produces two superimposed images whose time delay can closely match $|\Delta t_{\rm c}|$. The microlensing hypothesis is maximally favored ($\log_{10}\mathscr{B}^{\rm L}_{\rm U}\gg1$) for $\mathscr{M}_{\rm B}/\mathscr{M}_{\rm A}\gtrsim1$ and equal SNRs, increasing with $|\Delta t_{\rm c}|$. The inferred redshifted lens masses lie in the range $M_{\rm L}^z\sim10^2$--$10^{5}~{\rm M}_{\odot}$ with impact parameters $y\sim0.1$--$3~{\rm R_E}$. Overall, the inferred Bayes factor depends on relative chirp-mass ratios, relative loudness, difference in coalescence times, and also the absolute SNRs of the overlapping signals. Cumulatively, our results indicate that overlapping black-hole binaries with nearly equal chirp masses and comparable loudness are likely to be falsely identified as lensed. Such misidentifications are expected to become more common as detector sensitivities improve. While our study focuses on ground-based detectors using appropriate detectability thresholds, the findings naturally extend to next-generation GW observatories.
\end{abstract}

\maketitle

\section{Introduction}
The detection of gravitational waves from merging compact binary coalescences (CBCs) has become routine with the Advanced LIGO~\cite{LIGOScientific:2014pky} and Advanced Virgo~\cite{VIRGO:2014yos} detectors, culminating in 90 events in the third Gravitational-wave Transient Catalog (GWTC-3)~\cite{KAGRA:2021vkt}. The detection rate has increased rapidly, from around twice a month to about twice a week during the current fourth observational run (O4)~\cite{GraceDB} of the detectors, where the number of significant detection candidates has more than doubled the GWTC-3 events by the end of the second part of O4. With planned upgrades to second-generation detectors, including KAGRA~\cite{Somiya:2011np, Aso:2013eba, KAGRA:2018plz, KAGRA:2020tym}, and the addition of LIGO-India~\cite{LigoIndia, Unnikrishnan:2013qwa, Ajith:2024inj}, the field is assured for a surge in observed events. Future detectors with enhanced low-frequency sensitivity, such as Einstein Telescope (ET)~\cite{Punturo:2010zz, Hild:2010id} and Cosmic Explorer (CE)~\cite{Reitze:2019iox, LIGOScientific:2016wof, Regimbau:2016ike}, will enable detection of longer inspiral cycles~\cite{ET:2019dnz, Sathyaprakash:2012jk, Branchesi:2023mws}. This will increase the probability that two or more GW signals will temporally overlap in the detector band~\cite{Regimbau:2009rk, Samajdar:2021egv, Pizzati:2022apa, Relton:2021cax, Antonelli:2021vwg, Himemoto:2021ukb, Janquart:2022fzz, Speri:2022kaq, Niu:2024wdi, Wang:2025ckw}.

\begin{table*}
    \renewcommand{\arraystretch}{1.5}
    \centering
    \begin{tabularx}{\linewidth}{c c X c}
    \toprule
    \textbf{Type} & \textbf{Ref.} & \multicolumn{1}{c}{\textbf{Remarks}} & \textbf{Concerned Regime} \\
    \midrule
    PE & \cite{Relton:2021cax} & Studied biases from differences in merger times, SNRs, chirp masses, and initial phases. & $|\Delta t_{\rm c}| < 0.1~{\rm s}$ \\
    PE & \cite{Samajdar:2021egv} & Analyzed overlaps of BBH and BNS signals, studying effects of mass, and SNRs. & $\Delta t_{\rm c}=0,\pm2~{\rm s}$\\
    PE & \cite{Antonelli:2021vwg} & Examined metrics for PE biases on individually resolved sources from the presence of confusion noise and waveform inaccuracies. & $|\Delta t_{\rm c}| < 1~{\rm s}$ \\
    PE & \cite{Himemoto:2021ukb} & Investigated statistical impacts of overlapping GWs using Fisher matrix analysis. & $|\Delta t_{\rm c}| < 0.1~{\rm s}$ \\
    Search & \cite{Relton:2022whr} & Evaluated search efficiency for overlapping signals with modeled and unmodeled pipelines. & $\Delta t_{\rm c} < 1~{\rm s}$ \\
    PE & \cite{Pizzati:2022apa} & Explored overlaps of non-spinning binaries, varying coalescence times, phases, and SNRs. & $\Delta t_{\rm c} < 0.5~{\rm s}$ \\
    PE & \cite{Janquart:2022fzz} & Implemented joint PE and hierarchical subtraction for high-mass, medium-SNR events. & $|\Delta t_{\rm c}|=1~{\rm s}$ \\
    TGR & \cite{Hu:2022bji} & Investigated impacts of overlapping signals and waveform inaccuracies on tests of GR. & $|\Delta t_{\rm c}|<4~{\rm s}$ \\
    PE & \cite{Wang:2023ldq} & Combined Fisher matrix and Bayesian analysis to compute biases in inferred posteriors. & $|\Delta t_{\rm c}|<0.1~{\rm s}$ \\
    PE & \cite{Langendorff:2022fzq} & Applied machine learning to PE of overlapped BBH systems. & $|\Delta t_{\rm c}|\leq0.05~{\rm s}$ \\
    PE & \cite{Alvey:2023naa} & Performed sequential simulation-based joint PE for efficient inference. & $\Delta t_{\rm c} = 0.05, 0.2, 0.5~{\rm s}$ \\
    TGR & \cite{Dang:2023xkj} & Studied overlapping impacts on post-Newtonian coefficients in GR tests. & $\Delta t_{\rm c}\in0.1-\mathcal{O}(10)~{\rm s}$ \\
    PE & \cite{Johnson:2024foj} & Investigated the time-frequency overlap and the impact on PE studies over BNS populations and binary parameters. & $|\Delta t_{\rm c}|<0.1~{\rm s}$\\
    \bottomrule
    \end{tabularx}
    \caption{Overview of previous works on overlapping GWs (listed chronologically) focused on parameter estimation (PE), searches, or biases in tests of GR. The final column highlights the range of coalescence time differences between the lower and the higher SNR signals, $\Delta t_{\rm c}$, indicating regimes of interest for studying overlapping signals.}
    \label{tab:review}
\end{table*}

Table~\ref{tab:review} summarises prior investigations of overlapping CBC signals, including their impact on search efficiency, PE biases, and tests of general relativity (GR). We further highlight the degree of overlap in each study by reporting the difference in coalescence times, $\Delta t_{\rm c}$, between the lower‑ and the higher‑SNR signal.

These studies show that parameter biases depend on signal-to-noise ratios (SNRs), chirp masses, and $\Delta t_{\rm c}$, with the largest biases typically arising for small time separations and similar binary black hole (BBH) intrinsic parameters~\cite{Relton:2021cax, Samajdar:2021egv, Himemoto:2021ukb, Antonelli:2021vwg, Pizzati:2022apa}. For instance, the inferred chirp masses can be biased by $20$--$30\%$ in the worst cases, especially for small merger time differences~\cite{Relton:2021cax, Himemoto:2021ukb}. The Fisher matrix analyses in Ref.~\cite{Pizzati:2022apa} and Ref.~\cite{Himemoto:2021ukb} conclude that biases are typically $10\%$--$30\%$ of the statistical errors for closely timed BBHs, and are more pronounced than for binary neutron star (BNS) mergers. Mitigation strategies such as hierarchical subtraction and joint parameter estimation~\cite{Janquart:2022fzz, Hu:2025vlp} have proven more reliable, though the latter is computationally expensive over the full $30$-dimensional parameter space of two BBHs. More recently, Ref.~\cite{Baka:2025yqx} have proposed using prior-informed Fisher matrices, overlap of time-frequency tracks to predict the magnitude of biases, as a systematic choice between hierarchical subtraction and joint parameter estimation depending on the expected bias and computational constraints. 

Search studies demonstrate that matched-filter searches~\cite{Relton:2022whr} and burst pipelines~\cite{Klimenko:2015ypf,Klimenko:2021ypf, Mishra:2024zzs} remain effective for $|\Delta t_{\rm c}| \geq 1~{\rm s}$, while BNSs overlap analyses~\cite{Johnson:2024foj} found that $91\%$ of the overlaps occur at low frequencies ($\lesssim 5$~Hz) where the inspiral evolution is slow. Overlapping signals have also been shown to bias tests of GR, particularly in post-Newtonian coefficients~\cite{Hu:2022bji, Dang:2023xkj, Gupta:2024gun}.

\begin{figure}[htb!]
     \centering
     \includegraphics[keepaspectratio, width=0.485\textwidth]{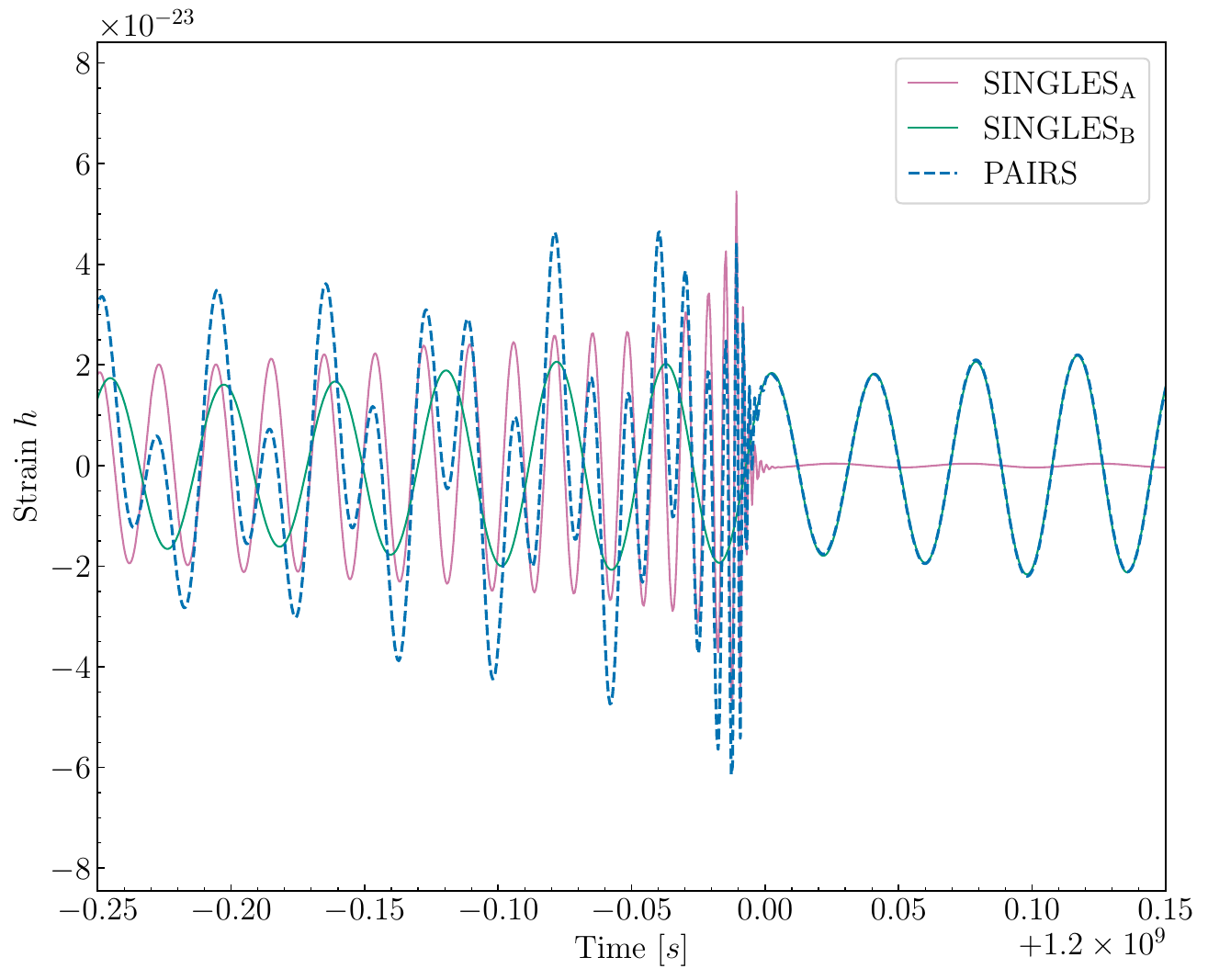}
     \includegraphics[keepaspectratio, width=0.485\textwidth]{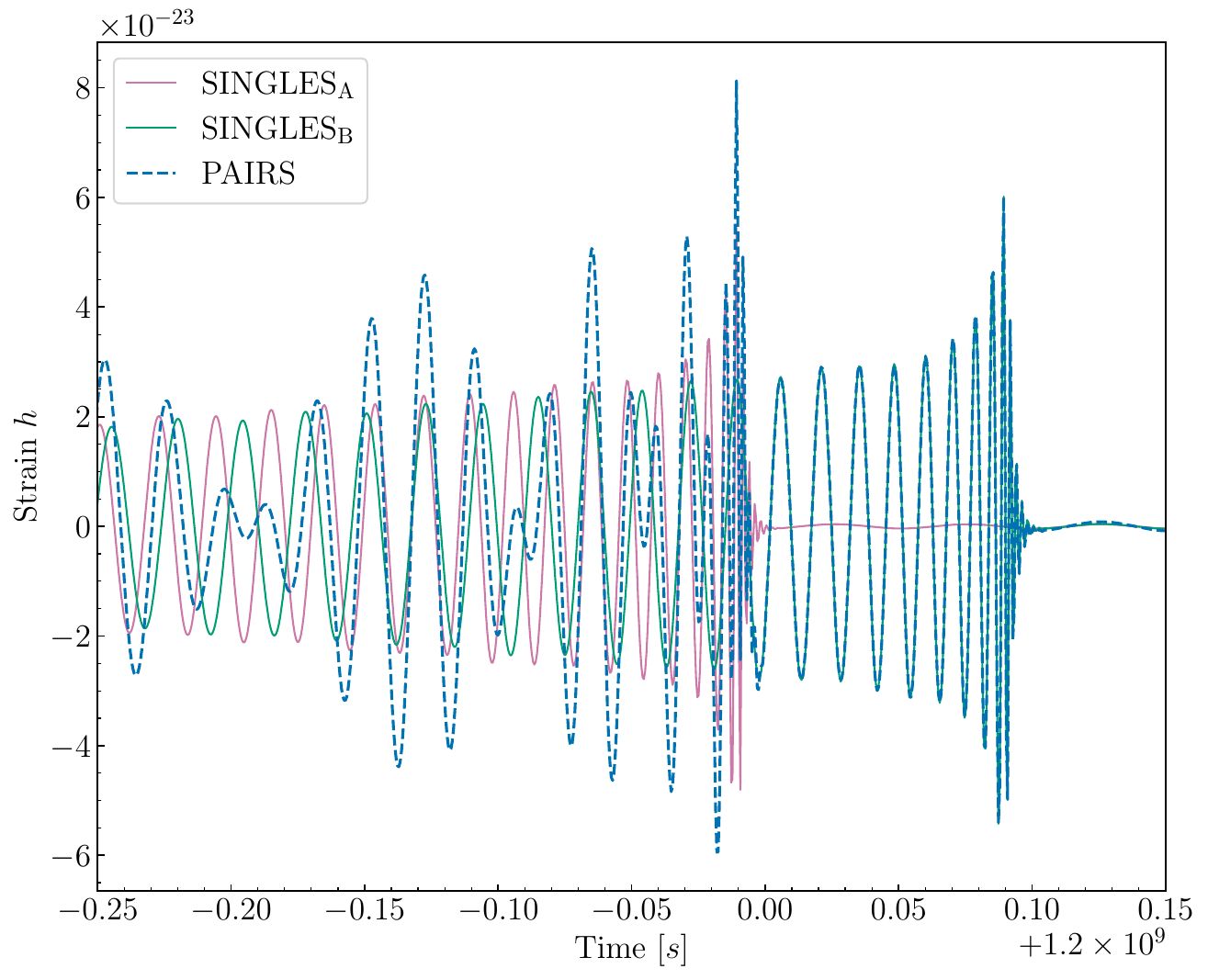}
     \caption{Representation of two BBH signals ($\rm{SINGLES_A}$ and $\rm{SINGLES_B}$) producing an overlapping strain (termed \textsc{PAIRS}). \textit{Top:} Different SNRs: The primary signal (termed $\rm{SINGLES_A}$, with $\mathrm{SNR}=30$) resembles the \textsc{GW150914}-like merger~\cite{LIGOScientific:2016vlm, LIGOScientific:2016lio, LIGOScientific:2018mvr}, and the secondary signal (termed $\rm{SINGLES_B}$, with $\mathrm{SNR}=15$) resembles the \textsc{GW170814}-like merger~\cite{LIGOScientific:2017ycc, LIGOScientific:2018mvr}, which coalesces $\Delta t_{\rm c}=0.15~{\rm s}$ after the merger of the primary signal. The louder signal dominates until its merger-ringdown phase, after which the quieter one becomes visible.
     \textit{Bottom:} Similar SNRs: both signals resemble the \textsc{GW150914}-like merger, producing a more strongly modulated waveform with no single signal dominating throughout, with $\Delta t_{\rm c}=0.1~{\rm s}$.}
     \label{fig:overlappedsignals}
\end{figure}

While these studies explored the impact of overlapping signals on searches and PE, none have examined the potential for degeneracies with other atypical physical effects, such as gravitational lensing and orbital eccentricity. Such a degeneracy could arise because overlapping CBC signals can produce amplitude and phase modulations resembling the frequency-dependent magnifications of \emph{microlensing}, the phase shifts of certain \emph{strongly lensed} images, and the modulation due to non-zero eccentricities. Fig.~\ref{fig:overlappedsignals} illustrates two examples of overlapping BBH signals with different and similar SNRs, showing how their superposition can lead to complex waveform morphologies.

This work investigates, for the first time, the degeneracy between overlapping CBC signals and gravitationally lensed signals in the context of ground-based detectors. By overlapping, we refer to two distinct (quasicircular, unlensed) BBH signals whose coalescence times differ by $|\Delta t_{\rm c}| < 0.1~{\rm s}$, with varied chirp mass ratios, SNRs, and temporal overlap. 

\begin{figure*}[htb!]
     \centering
     \includegraphics[keepaspectratio, width=1\textwidth]{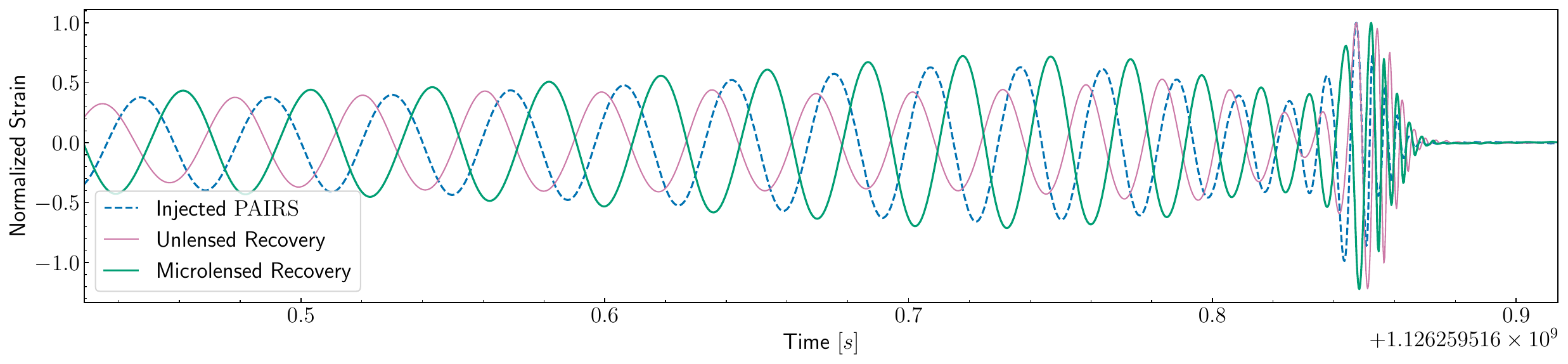}
     \includegraphics[keepaspectratio, width=1\textwidth]{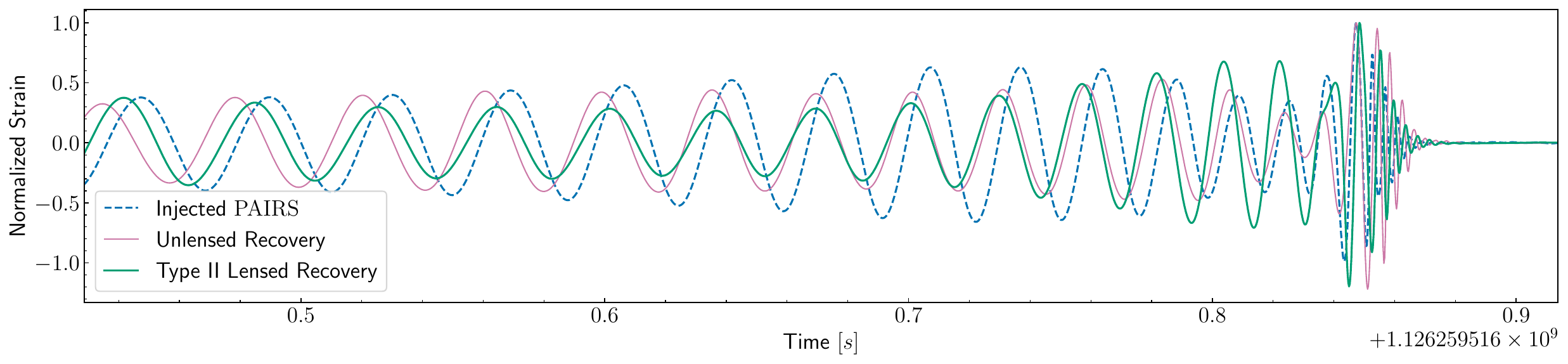}
     \caption{Illustration of waveform morphology similarities between an overlapping signal and its best-fit lensed counterparts. The injected overlapping waveform at the detector and the corresponding best-fit normalized lensed waveforms obtained from the inferred values from PE are shown. The normalized strain is defined as the ratio of strain $h$ to the maximum amplitude $|h|$. \textit{Top:} Contains an injected overlapping waveform (termed \textsc{PAIRS}) from a combination of two GW signals ($\rm{SINGLES_A}$ and $\rm{SINGLES_B}$) with relative parameters of chirp mass ratio, $\mathscr{M}_{\rm B}/\mathscr{M}_{\rm A} = 0.5$, SNR ratio, $\mathrm{SNR}_{\rm B}/\mathrm{SNR}_{\rm A} = 1$, and coalescence time difference, $\Delta t_{\rm c} = 0.02~{\rm s}$, and its corresponding inferred Unlensed and Microlensed waveforms. \textit{Bottom:} Contains the same overlapping strain as above, but with the inferred Type-II lensed waveform. Notably, the microlensed model provides a visibly better match to the modulated structure of the overlapping signal, especially near the merger phase.}
     \label{fig:mlpewf}
\end{figure*}

We focus on two lensing scenarios: (i) \emph{Microlensing} in the wave-optics regime, where the GW wavelength is comparable to the Schwarzschild radius of an intervening compact lens, such as a stellar-mass black hole~\cite{Nakamura:1999uwi, Takahashi:2003ix}, and (ii) \emph{Type-II Strongly Lensed images}, produced by galaxy- or cluster-scale lenses and differing from type-I images by a relative phase shift of $-\pi/2$~\cite{Dai:2017huk}. Fig.~\ref{fig:mlpewf} compares an injected overlapping GW signal to its best-fit lensed counterparts, illustrating how these models can reproduce the overlap-induced modulations.

Since performing joint PE for overlapping signals is computationally far more expensive than analyzing them under the microlensing or type-II lensing hypotheses, we adopt an approach in which we only \textit{inject} overlapping signals and use lensing as the inference model for computing evidences (i.e., marginalized likelihoods). By \textit{injection}, we refer to simulated observations of two overlapping GW signals in the detectors, termed as $\rm{SINGLES_A}$ and $\rm{SINGLES_B}$, with $\rm{SINGLES_A}$ having the higher SNR. Specifically, we consider only zero-noise injections to isolate the effects arising purely from overlapping signals; noise-systematic effects are left to future work. Using the three-detector network of Hanford, Livingston, and Virgo at A+ design sensitivity, we compare and evaluate the extent to which the two lensing models can mimic overlapping GW signals. We use fitting factor (FF) calculations and PE techniques to analyze the overlapping signals and identify regions in their parameter space, characterized by chirp mass ratios, SNR ratios, and temporal overlap, that can potentially be misinterpreted as lensing effects.

The paper is organized as follows: Sec.~\ref{sec:lensing} outlines the basics of gravitational lensing, Sec.~\ref{sec:methods} describes the methods for analyzing overlapping signals, Sec.~\ref{sec:setups} explains the analysis setup, Sec.~\ref{sec:results} presents the results, and Sec.~\ref{sec:conclusions} concludes with an outlook.

\section{Gravitational Lensing}
\label{sec:lensing}

Gravitational lensing of a GW signal occurs when it passes through the gravitational potential of an intervening astrophysical object, such as a galaxy, galaxy cluster, star, or black hole. The effect depends on the mass distribution of the lens and the source-lens-observer alignment, but in all cases, the lensed waveform can be expressed as,
%-------------
\begin{equation}
\tilde{h}_{\rm L}(f; \lambda) = \tilde{h}(f; \lambda) \, F(f,\Theta)\,,
\label{eq:lens_waveform}
\end{equation}
%-------------
where $\tilde{h}(f; \lambda)$ is the unlensed waveform in the frequency domain, $\lambda$ denotes the intrinsic and extrinsic binary source parameters (masses, spins, sky location, luminosity distance, inclination, polarisation, coalescence time, and phase), and $\Theta$ describes the lens parameters. For an isolated point-mass lens, $\Theta = \{M_{\rm L}^z,\, y\}$, where $M_{\rm L}^z \equiv M_{\rm L}(1+z_{\rm L})$ is the redshifted lens mass with $z_{\rm L}$ being the redshift of the lens, and $y$ is the source–lens angular separation in units of the Einstein radius, ${\rm R_E}$.

The complex, frequency-dependent amplification factor $F(f,\Theta)$ encodes the lensing effect. For a given lens profile, it can be computed using the diffraction integral~\cite{Takahashi:2003ix},
%-------------
\begin{equation}
F(w, y) = \frac{w}{2\pi i} \int d^{2}x\, \exp\!\left[iwT(x, y)\right]\,,
\label{eq:amplification}
\end{equation} 
%-------------
where $x$ and $y$ are the image- and source-plane coordinates in units of the lens scale $\xi_0$ (typically the Einstein radius, ${\rm R_E}$), and $T(x,y)$ is the dimensionless time-delay function. The dimensionless frequency is
%-------------
\begin{equation}
w = \frac{D_{\rm OS}}{D_{\rm OL}D_{\rm LS}} \xi_{0}^{2} (1 + z_{\rm L}) 2\pi f\,,
\label{eq:frequency}
\end{equation} 
%-------------
with $D_{\rm XY}$ the angular-diameter distances between observer (O), lens (L), and source (S). The value of $w$ determines the relevant lensing regime.

\paragraph*{Strong lensing (geometric optics):} 
When $w \gg 1$ for the frequencies of interest, only stationary points of $T(x, y)$ contribute significantly to Eq.~\eqref{eq:amplification}. In the geometric optics limit, we expect a combination of multiple images, characterized by the relative phase shift, time delays, and magnification. The amplification factor for image $j$ reduces to~\cite{Ezquiaga:2020gdt, Dai:2017huk},
%-------------
\begin{equation}
F_{j}(f) = |\mu_{j}|^{1/2} \exp\!\left[2\pi i f t_{j} - i\pi n_{j}\,\mathrm{sgn}(f)\right]\,.
\label{eq:geometric-amplification}
\end{equation}
%-------------
Here $\mu_{j}$ is the magnification, $t_{j}$ is the arrival-time delay due to path length and Shapiro delay, and $n_{j}$ is the Morse index determined by the image type: $n_{j}=0$ (Type-I, minima), $n_{j}=1/2$ (Type-II, saddle points), or $n_{j}=1$ (Type-III, maxima), corresponding to phase shifts of $0$, $-\pi/2$, and $-\pi$, respectively. In this work, we model \emph{Type-II} images following previous studies~\cite{Vijaykumar:2022dlp, Wright:2024mco}, in which the $-\pi/2$ phase shift imprints frequency-dependent distortions across the waveform.

\paragraph*{Microlensing (wave optics):}
When $w \sim \mathcal{O}(1)$, the GW wavelength is comparable to the Schwarzschild radius of the lens, and the geometric-optics approximation fails. The full diffraction integral in Eq.~\eqref{eq:amplification} must be evaluated. For an isolated point-mass lens, $F(w, y)$ admits an analytical expression in terms of a hypergeometric expansion~\cite{Takahashi:2003ix, Nakamura:1999uwi, LIGOScientific:2023bwz, Chan:2025wgz},
%-------------
\begin{multline}
    F(w, y) = \exp\left\{\frac{\pi\,w}{4} + \frac{i\omega}{2}\left[\ln\left(\frac{w}{2}\right) - 2\phi_{\rm m}\left(y\right)\right]\right\} \\ \times\,\Gamma\left(1-\frac{iw}{2}\right)
    {}_{1}\mathsf{F}_{1}\left[\frac{iw}{2},1;\frac{i\omega y^2}{2}\right]\,,
    \label{eq:wave-amplification}
\end{multline}
%-------------
as a complex, frequency-dependent magnification that produces characteristic oscillatory modulations in the amplitude and phase of $\tilde{h}_{\rm L}(f;\lambda)$. The dimensionless frequency is characterized by $w=8\pi GM_{\rm L}^{z}\,f/c^3$, and the frequency-independent phase $\phi_{\rm m}(y)=\frac{1}{2}\left(x_{\rm m}-y\right)^2-\ln x_{\rm m}$ describes the effective potential in the presence of the lens, with $x_{\rm m}=\frac{1}{2}y+\frac{1}{2}\sqrt{y^2+4}$.

Using the methods outlined in Ref.~\cite{Basak:2021ten} and Ref.~\cite{Mishra:2023ddt}, we generate microlensed signals with an isolated point lens model~\cite{Schneider:1992bmb}. This simple yet effective lensing profile enables detailed parameter estimation and mismatch analysis for overlapping signals. Previous studies~\cite{Janquart:2023mvf, Wright:2024mco} have extensively employed this model in microlensing searches due to its computational efficiency.

This study uses Type-II strong-lensing and point-mass microlensing models as inference templates for overlapping GW signals. Both models can produce characteristic amplitude and phase modulations; a $-\pi/2$ phase shift for Type-II images in the geometric-optics limit, and oscillatory patterns in the wave-optics regime for microlensing. These patterns may be similar to overlapping signatures, particularly in the geometric optics limit with small time delay configurations, creating a potential for degeneracy. These models have been employed in GW lensing studies~\cite{Meena:2019ate, Basak:2021ten, Mishra:2021xzz, Vijaykumar:2022dlp, Janquart:2023mvf, LIGOScientific:2023bwz, Mishra:2023ddt, Mishra:2023vzo, Wright:2024mco} and are therefore well-suited for the PE and FF analyses presented in this study. However, it is worth noting that model-independent approaches for detecting lensed signals also exist~\cite{Liu:2023ikc, Chakraborty:2025maj}, although they are not utilized in this study.

\section{Descriptions of Methods}
\label{sec:methods}

\subsection{Parameter Estimation Techniques}
\label{subsec:pe}

This work uses Bayesian parameter estimation to test whether overlapping GW signals are better inferred with lensed or unlensed templates and to quantify the resulting parameter biases. Given data $d$ and hypothesis $\mathcal{H}$, the posterior for parameters $\theta$ is
%-------------
\begin{equation}
    p(\theta | d, \mathcal{H}) = \frac{\mathcal{L}(d | \theta, \mathcal{H})\,\pi(\theta | \mathcal{H})}{\mathcal{Z}(d | \mathcal{H})}\,,
    \label{eq:BayesTheorem}
\end{equation} 
%-------------
where $\pi(\theta | \mathcal{H})$ is the prior, $\mathcal{L}(d | \theta, \mathcal{H})$ is the likelihood, and $\mathcal{Z}(d | \mathcal{H})$ is the evidence, which is the marginalised likelihood.

Assuming stationary Gaussian noise, the likelihood is
%-------------
\begin{equation}
    \mathcal{L}(d | \theta, \mathcal{H}) \propto \exp\!\left[ -\frac{1}{2} \langle d - h(\theta) | d - h(\theta) \rangle \right]\,,
    \label{eq:likelihood}
\end{equation} 
%-------------
with the noise-weighted inner product
%-------------
\begin{equation}
    \langle a | b \rangle = 4\,\Re \int_{f_{\min}}^{f_{\max}}  \frac{\tilde{a}(f)\,\tilde{b}^{*}(f)}{S_{n}(f)}\, df\,.
    \label{eq:innerproduct}
\end{equation}
%-------------
Here, $S_n(f)$ is the power spectral density (PSD) of the detector, and $f_{\min}$ and $f_{\max}$ are the analysis frequency bounds\footnote{Typically, $f_{\min}$ corresponds to the lowest frequency where the detector noise can be approximated as stationary, while $f_{\max}$ denotes the Nyquist frequency.}. We use the target PSDs for the O4 observing run of the Advanced LIGO and Virgo detectors\footnote{For the LIGO detectors, we use the PSD provided \href{https://dcc.ligo.org/public/0165/T2000012/002/aligo_O4high.txt}{here}, and for Virgo, we use the PSD available \href{https://dcc.ligo.org/public/0165/T2000012/002/avirgo_O4high_NEW.txt}{here}.}~\citep{KAGRA:2013rdx}.

The evidence enables model selection under the Bayesian framework. Comparing a lensed hypothesis ($\mathcal{H}_{\rm L}$) with an unlensed one ($\mathcal{H}_{\rm U}$) yields the Bayes factor
%-------------
\begin{equation}
    \log_{10}\mathscr{B}^\mathrm{L}_{\rm U} = \log_{10}\mathcal{Z}_{\rm L} - \log_{10}\mathcal{Z}_{\rm U}\,.
    \label{eq:logbf}
\end{equation}
%-------------

Since lensing effects are applied on top of the standard quasicircular unlensed waveform, $\pi_{\rm U}$ is a subset of $\pi_{\rm L}$. The PE results in Sec.~\ref{sec:results} quantify both the inference accuracy and the statistical preference between models.

\begin{table*}
    \centering
    \renewcommand{\arraystretch}{1}
    \begin{tabular}{c c c}
    \toprule
    \textbf{Parameter} & \textbf{Population Model} & \textbf{PE Priors} \\
    \midrule
    \multicolumn{3}{c}{\emph{BBH Parameters}} \\
    \midrule
    $\mathscr{M}$: Chirp Mass & -- & $\mathcal{U}(15, 50)~{\rm M}_{\odot}$ \\
    $q$: Mass ratio & -- & $\mathcal{U}(0.05,1)$ \\
    $m_1,m_2$: Component Masses & \textsc{PowerLaw + Peak}~\cite{LIGOScientific:2021djp}: $[5, 100]~{\rm M}_{\odot}$ & Constraint: $[3.5,130]~{\rm M}_{\odot}$ \\
    $z$: Redshift & \textsc{PowerLaw}~\cite{Fishbach:2018edt} & -- \\
    $d_L$: Luminosity distance & Interpreted from $z$ & \textsc{PowerLaw}$_{\alpha=2}(50,5000)~\rm{Mpc}$ \\
    $a_{1/2}$: Spin amplitude 1/2 & Beta distribution (i.i.d\footnote{i.i.d refers to independent and identically distributed~\cite{Wysocki:2018mpo}.}) & $\mathcal{U}(0, 1)$ \\
    $\theta_{1/2}$: Tilt angle 1/2 & Truncated normal in cosine (i.i.d) & Uniform in $\sin\theta$ \\
    $\phi_{12}$: Spin vector azimuthal angle & $\mathcal{U}(0, 2\pi)$ & $\mathcal{U}(0, 2\pi)$ \\
    $\iota$: Inclination angle & $\mathcal{U}(0, \pi)$ & $\mathcal{U}(0, \pi)$ \\
    $\psi$: Wave polarisation & $\mathcal{U}(0, \pi)$ & $\mathcal{U}(0, \pi)$ \\
    $\Phi_{\rm c}$: Phase of coalescence & $\mathcal{U}(0, 2\pi)$ & $\mathcal{U}(0, 2\pi)$ \\
    $\alpha$: Right ascension & $\mathcal{U}(0, 2\pi)$ & $\mathcal{U}(0, 2\pi)$ \\
    $\delta$: Declination & Uniform in $\cos\delta$ & Uniform in $\cos\delta$ \\
    \midrule
    \multicolumn{3}{c}{\emph{Type-II Strong Lensing Parameter}} \\
    \midrule
    $n_j$: Morse Phase & -- & Fixed: $\delta(n_j - 0.5)$ \\ 
    & & Varying: $\mathcal{U}(0, 1)$ \\
    \midrule
    \multicolumn{3}{c}{\emph{Microlensing Parameters}} \\
    \midrule
    $y$: Impact Parameter & -- & \textsc{PowerLaw}$_{\alpha=2}(0.01,5)~{\rm R_E}$\\
    $M^z_{\rm L}$: Redshifted Lens Mass & -- & Log-Uniform in $[0.1, 10^5]~{\rm M}_{\odot}$\\ 
    \bottomrule
    \end{tabular}
    \caption{Overview of the functions used to generate the BBH populations for $\rm{SINGLES_A}$ and $\rm{SINGLES_B}$, along with the priors employed in parameter estimation. Additional priors specific to microlensed and Type-II analyses are also included.}
    \label{tab:params}
\end{table*}

\subsection{Fitting Factor Analysis}
\label{subsec:ff}

To complement PE inferences, which are computationally expensive, we use fitting factor studies~\cite{Owen:1995tm, Owen:1998dk} as fast diagnostics of waveform degeneracy between overlapping signals and lensed templates.

The overlap between two waveforms is
%-------------
\begin{equation}
    \mathcal{O}[h_1, h_2] = \frac{\langle h_1 | h_2 \rangle}{\sqrt{\langle h_1 | h_1 \rangle \langle h_2 | h_2 \rangle}}\,,
    \label{eq:overlap}
\end{equation}
%-------------
and the match $\mathcal{M}[h_1, h_2]$ is the maximum overlap over coalescence time $t_{\rm c}$ and phase $\Phi_{\rm c}$.

The fitting factor is
%-------------
\begin{equation}
    \mathscr{F} = \max_{\theta}\mathcal{M}[h_{\rm s}, h_{\rm T}(\theta)]\,,
    \label{eq:ffmatch}
\end{equation}
%-------------
where $h_{\rm s}=\sum_{i\in\{{\rm A},{\rm B}\}} h_i$ is the overlapping GW signal and $h_{\rm T}(\theta)$ is the template. 

Optimization over intrinsic parameters is performed using Differential Evolution (DE), which ensures robustness across parameter space, unlike the Nelder-Mead (NM) method~\cite{Lagarias:1998iqa, Jaranowski:2005hz, Ajith:2007kx, Ajith:2007xh, CalderonBustillo:2015lrt, Basak:2021ten, Mishra:2023ddt, Ajith:2024inj}, which is biased by the initial data. Extrinsic parameters other than $t_{\rm c}$ and $\Phi_{\rm c}$ are excluded, as they do not significantly alter the intrinsic phase evolution of the waveform in the dominant mode and appear only as overall amplitude scalings or constant phase shifts arising from detector projection.

The fitting factor can be related to the recovered matched-filter SNR~\cite{Ajith:2012mn, Usman:2015kfa},
%-------------
\begin{equation}
    \langle\rho(\theta)\rangle=\mathscr{F}\cdot \rho_{\rm opt}\,,
    \label{eq:snr_ff}
\end{equation} 
%-------------
where $\rho_{\rm opt}=\langle h| h\rangle=\max\limits_{\theta}\langle \rho(\theta)\rangle$is the optimal SNR.

In the high-SNR limit, this yields an approximate relation between the FF and the Bayes factor~\cite{Cornish:2011ys}, through a Laplace approximation~\cite{AzevedoFilho:2013lap},
%-------------
\begin{equation}
    \ln\mathscr{B}^{\rm L}_{\rm U} \approx \frac{1}{2}\rho_{\rm opt}^2\left(\mathscr{F}_{\rm L}^2-\mathscr{F}_{\rm U}^2\right)\,,
    \label{eq:logbf_ff}
\end{equation} 
%-------------
assuming equal noise evidence and neglecting Occam factors. We compute network fitting factors via a quadrature sum across detectors, analogous to the network SNR. High FF values ($\mathscr{F} \gtrsim 1$) correspond to $\sim 90\%$ detection efficiency~\cite{Cornish:2011ys, Sampson:2013wia, GilChoi:2022waq}, ensuring minimal SNR loss.

By combining PE and FF analyses, we can assess the parameter space of overlapping signals that is degenerate with lensing.

\section{Setup of the Analyses} 
\label{sec:setups}

Our study investigates pairs of overlapping BBH signals, analysed with both Type-II strong-lensing and microlensing waveform models. We generate three injection sets: two with individual, non-overlapping BBHs ($\rm{SINGLES_A}$ and $\rm{SINGLES_B}$) and one containing overlapping pairs (\textsc{PAIRS}). Waveforms are generated with the \texttt{IMRPhenomXPHM} approximant~\cite{Hannam:2013oca,Husa:2015iqa,Khan:2015jqa,Pratten:2020ceb,Ramos-Buades:2023ehm} from the \texttt{LALSuite} package~\cite{lalsuite,swiglal}, setting spins to zero unless otherwise stated.

\paragraph*{Relative parameter space:}
We characterise an overlapping pair by the chirp-mass ratio $\mathscr{M}_{\rm B}/\mathscr{M}_{\rm A}$, the network-SNR ratio $\mathrm{SNR}_{\rm B}/\mathrm{SNR}_{\rm A}$, and the coalescence-time difference $\Delta t_{\rm c} = t_{\rm c}^{\rm B} - t_{\rm c}^{\rm A}$. These quantities control the combined waveform morphology: $\mathscr{M}$ influences the signal duration, the SNR ratio sets the relative amplitude, and $\Delta t_{\rm c}$ governs the degree of overlap~\cite{Regimbau:2009rk, Samajdar:2021egv, Pizzati:2022apa, Relton:2021cax, Antonelli:2021vwg, Himemoto:2021ukb, Janquart:2022fzz, Speri:2022kaq}. Without loss of generality, $\rm{SINGLES_A}$ is defined to have the higher SNR, since, if two different signals arrive within $0.1~{\rm s}$ window (trigger window considered for searches~\cite{Usman:2015kfa, Davies:2020tsx}), they are most likely to be analysed as a single event. In our preliminary analyses and through matched filtering studies~\cite{Relton:2021cax, Johnson:2024foj}, we have observed that the parameters of the signal with higher SNR are obtained.

We explore a $3$D grid of $60$ points: $\mathscr{M}_{\rm B}/\mathscr{M}_{\rm A} \in \{0.5, 1, 2\}$, $\mathrm{SNR}_{\rm B}/\mathrm{SNR}_{\rm A} \in \{0.5, 1\}$, and $\Delta t_{\rm c}$ logarithmically sampled in $[-0.1, 0.1]$\,s (10 samples in the strong-overlap regime). $\rm{SINGLES_A}$ parameters are based on \textsc{GW150914}~\cite{LIGOScientific:2016vlm,LIGOScientific:2016lio,LIGOScientific:2018mvr} and $\rm{SINGLES_B}$ on \textsc{GW170814}~\cite{LIGOScientific:2017ycc,LIGOScientific:2018mvr}, with luminosity distances scaled so that $\rm{SINGLES_A}$ has network optimal SNR $\rho_{\rm opt} = 30$. The mass ratios of $\rm{SINGLES_A}$ and $\rm{SINGLES_B}$ are consequently fixed to $0.81$ and $0.87$, respectively, which are nearly identical. The arrival time differences are sampled to include negative values as a potential (false) identification of lensing, since the lens models considered for the study produce images with a positive time delay and demagnification and a corresponding phase shift. For $\Delta t_{\rm c}<0$, we have the case where the louder signal coalesces after the weaker, which results in a magnified image.

We further extend the analysis to a BBH population samples of $\mathcal{O}(5000)$ signals, drawn from GWTC-3–like distributions. The BBH waveforms were generated using the \texttt{IMRPhenomXPHM} waveform approximant, with the parameter distributions in Table~\ref{tab:params}, sampled with the \texttt{gwpopulation} package~\cite{Talbot:2019okv}. A threshold network SNR of 12 ensured the inclusion of detectable signals, with $\Delta t_{\rm c}$ logarithmically sampled in $[-0.1, 0.1]~\mathrm{s}$.

\paragraph*{Detector configuration and noise model:}
All injections use the HLV network (Advanced LIGO Hanford and Livingston at design sensitivity~\cite{LIGOScientific:2014pky}, Advanced Virgo~\cite{VIRGO:2014yos}), analyzed under zero-noise conditions to isolate waveform-systematic effects from noise fluctuations.

\paragraph*{Analysis pipeline:}
Bayesian PE is performed on the $60$ signals with \texttt{BILBY}~\cite{Ashton:2018jfp,Smith:2019ucc,Romero-Shaw:2020owr} using the \texttt{Dynesty} sampler~\cite{Speagle:2019ivv}. Table~\ref{tab:params} summarizes the injection distributions and PE priors, including lensing-specific ones. Type-II strong-lensing templates incorporate the Morse index $n_j$ as a $-\pi/2$ phase shift, and further vary the Morse index as a free parameter. Microlensed templates include the frequency-dependent amplification factor computed via the \texttt{gwmat} package\footnote{\url{https://git.ligo.org/anuj.mishra/gwmat/}}~\cite{Mishra:2023ddt}. For each overlapping pair, analyses are repeated with unlensed templates to assess degeneracies between overlap effects and lensing.

Fitting factors are computed following Eq.~\eqref{eq:ffmatch} using \texttt{PyCBC}~\cite{pycbc} with a DE optimization algorithm~\cite{Storn:1997uea, Virtanen:2019joe}. Intrinsic parameters to be maximized over, cover $\{\mathscr{M}, \eta\}$ for unlensed signals, extended to include $\{n_j\}$ for Type-II and $\{M_{\rm L}^z, y\}$ for microlensing. The FF-derived Bayes factor is estimated via Eq.~\eqref{eq:logbf_ff} for the extended BBH population.

\section{Results and Discussions}
\label{sec:results}

We present results for overlapping GW transients using two complementary approaches:(i) Bayesian PE for a discrete set of cases, and (ii) FF studies over a broader BBH population. Both analyses are used to quantify parameter biases and identify regions of degeneracy when overlapping signals are inferred with lensing models.

\begin{figure*}[htb!]
     \centering
     \includegraphics[keepaspectratio, width=.33\textwidth]{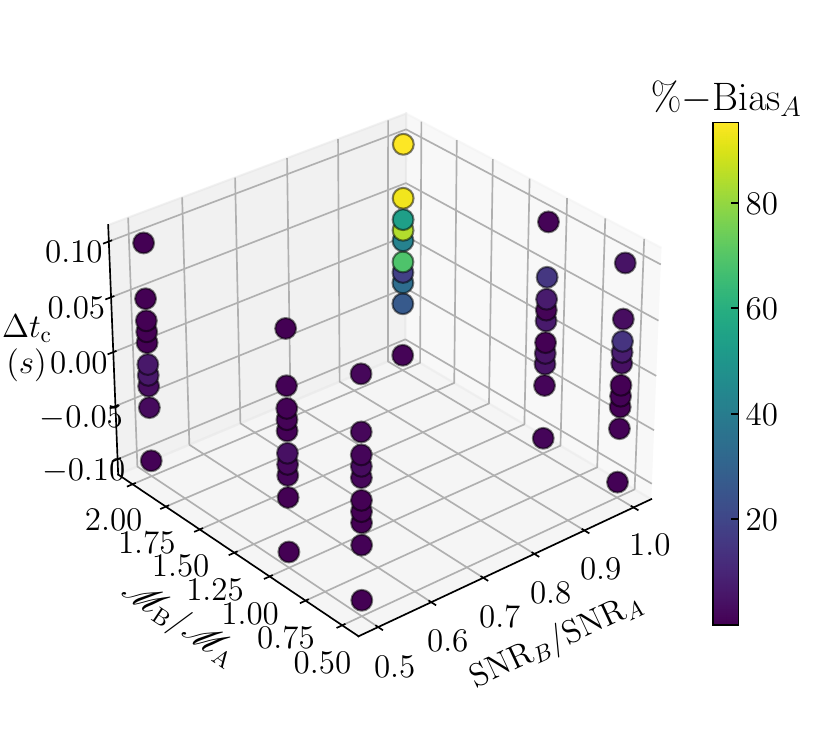}%
     \includegraphics[keepaspectratio, width=.33\textwidth]{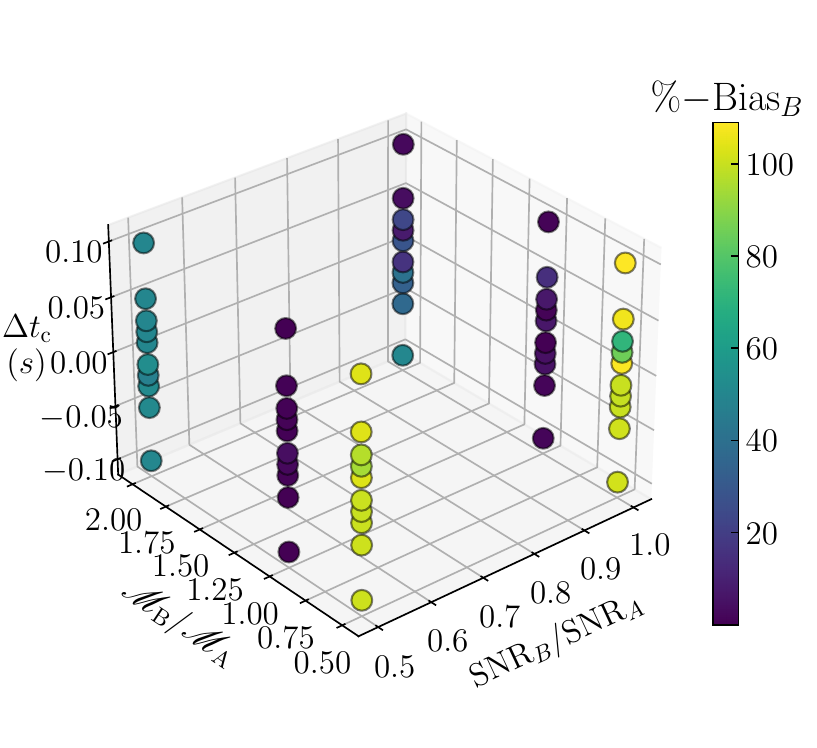}%
     \includegraphics[keepaspectratio, width=.33\textwidth]{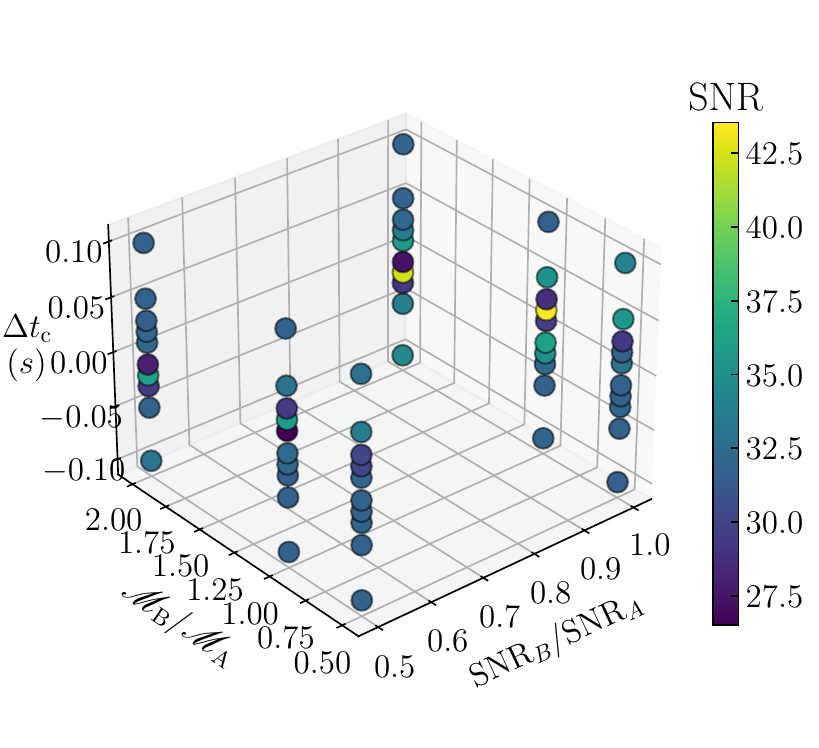}
     \caption{Inferred relative biases in the inferred chirp mass and the recovered SNR from PE on injected overlapping signals, assuming an unlensed quasicircular signal model. The chirp mass bias for each signal is defined as the relative difference between the inferred chirp mass and the mass of the injected signal. Variations in the recovered SNR reflect both the differences in the true injected SNR and the mismatches between the injections and the unlensed templates.}
     \label{fig:singlespe}
\end{figure*} 

\begin{figure*}[htb!]
     \centering
     \includegraphics[keepaspectratio, width=0.33\textwidth]{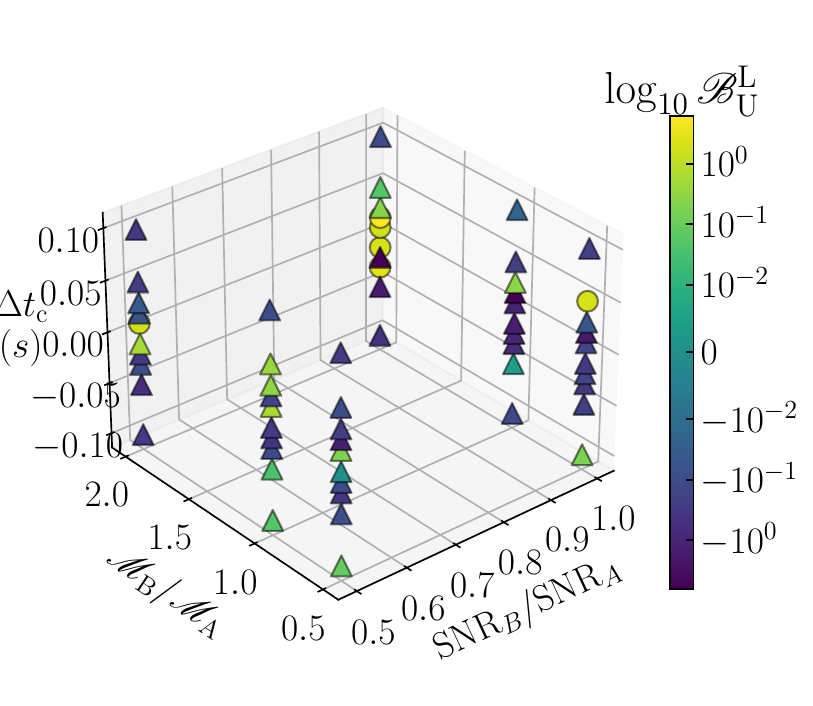}%
     \includegraphics[keepaspectratio, width=0.33\textwidth]{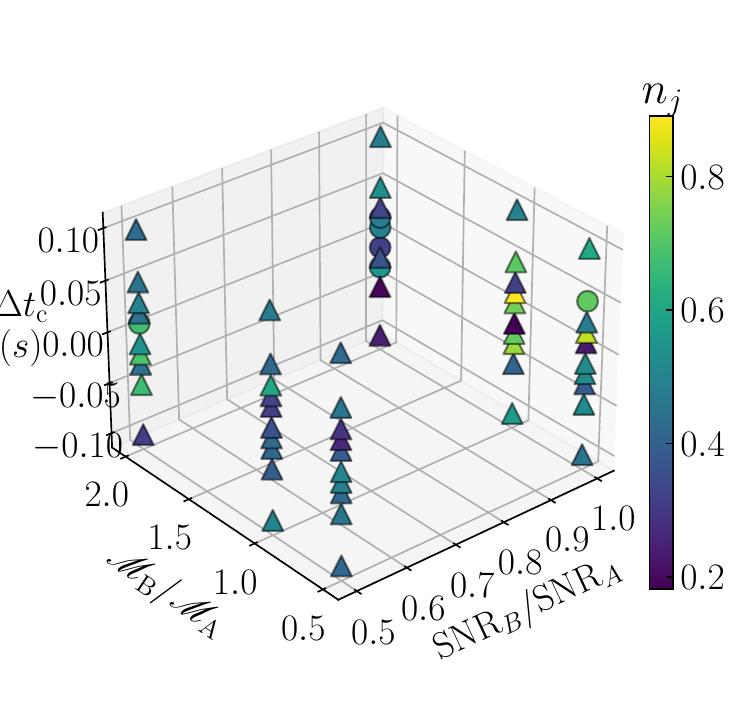}%
     \includegraphics[keepaspectratio, width=0.33\textwidth]{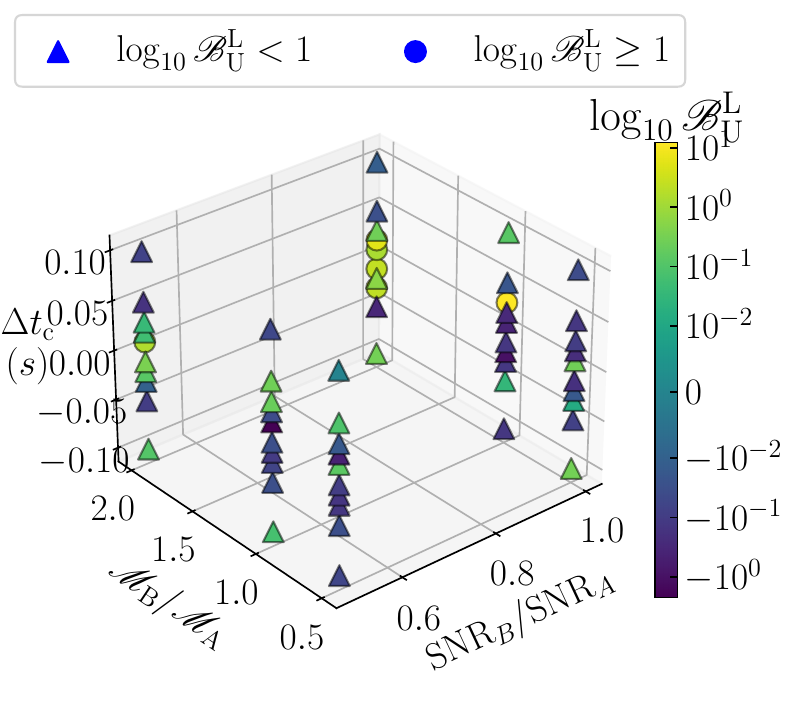}
     \caption{Results of fitting Type-II lensed templates to overlapping signals as a function of chirp mass ratio, SNR ratio, and coalescence time difference. \emph{Left:} Inferred Bayes factors for a Type-II image ($n_j=0.5$). \emph{Middle and Right:} Results when the Morse phase $n_j$ is a free parameter. Circular markers indicate $\log_{10}\mathscr{B}^{\rm L}_{\rm U} \geq 1$, corresponding to strong support for lensing.}
     \label{fig:slpeparams}
\end{figure*} 

\begin{figure}[htb!]
     \centering
     \includegraphics[keepaspectratio, width=.485\textwidth]{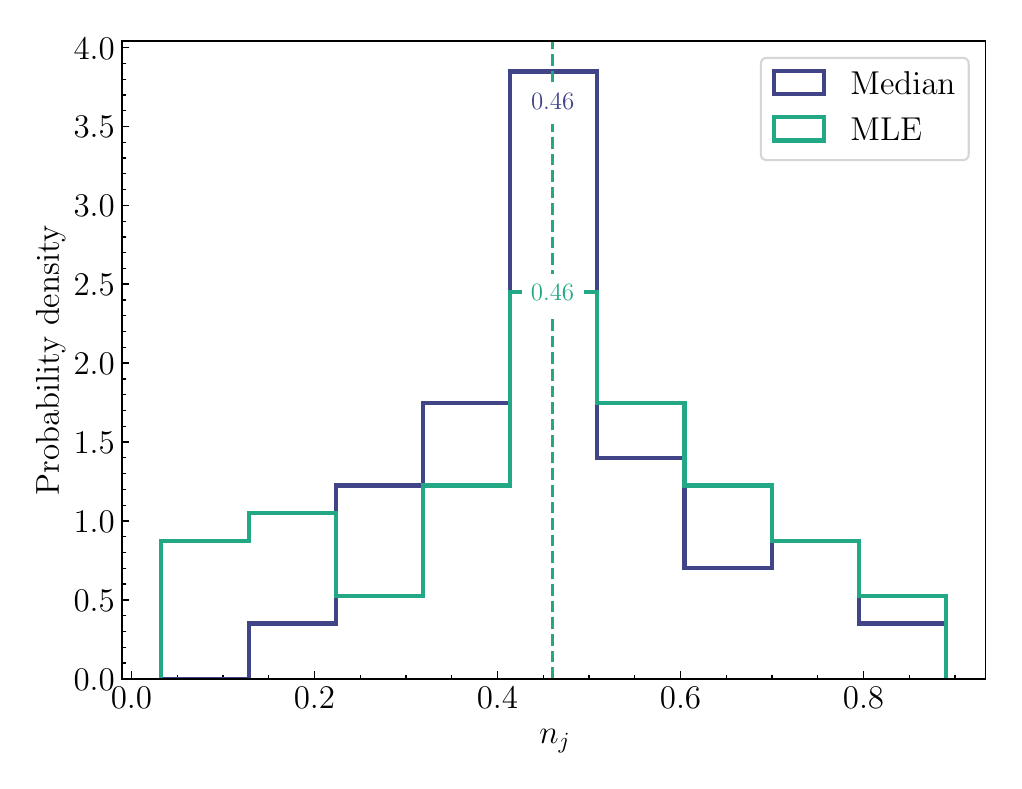}
     \caption{Inferred Morse index $n_j$ densities from parameter estimation under the Type-II lensing hypothesis for all $60$ injections. Posterior medians (\textit{purple}) and maximum likelihood estimate, MLE (\textit{green}) values show clustering near $n_j=0.5$, indicating a preference for a saddle-point (Type-II) image. Dashed lines denote the overall medians across the $60$ runs.}
     \label{fig:nslpe}
\end{figure} 

\begin{figure*}[htb!]
     \centering
     \includegraphics[keepaspectratio, width=0.33\textwidth]{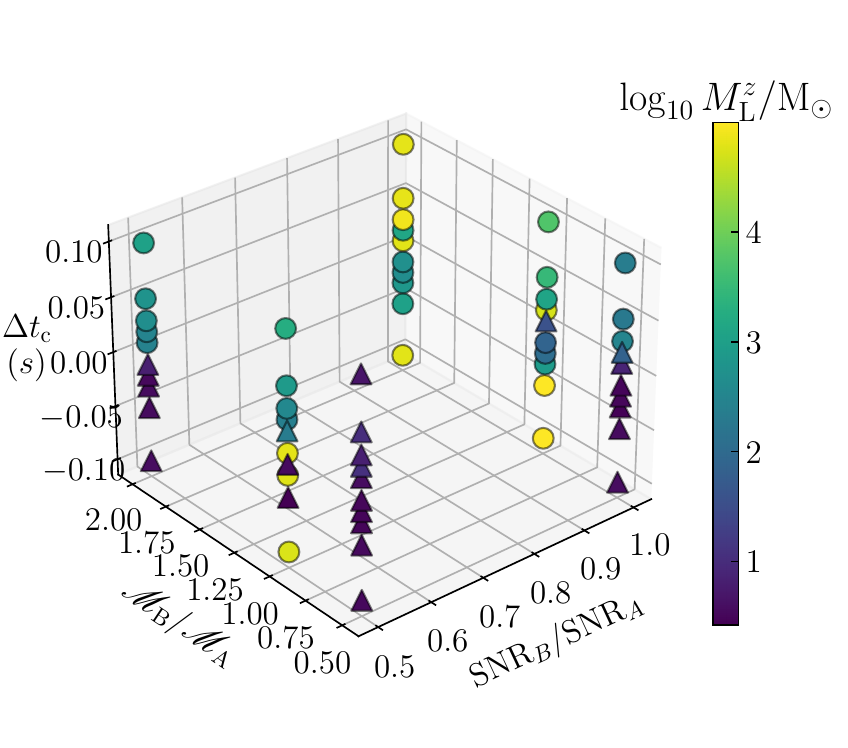}%
     \includegraphics[keepaspectratio, width=0.33\textwidth]{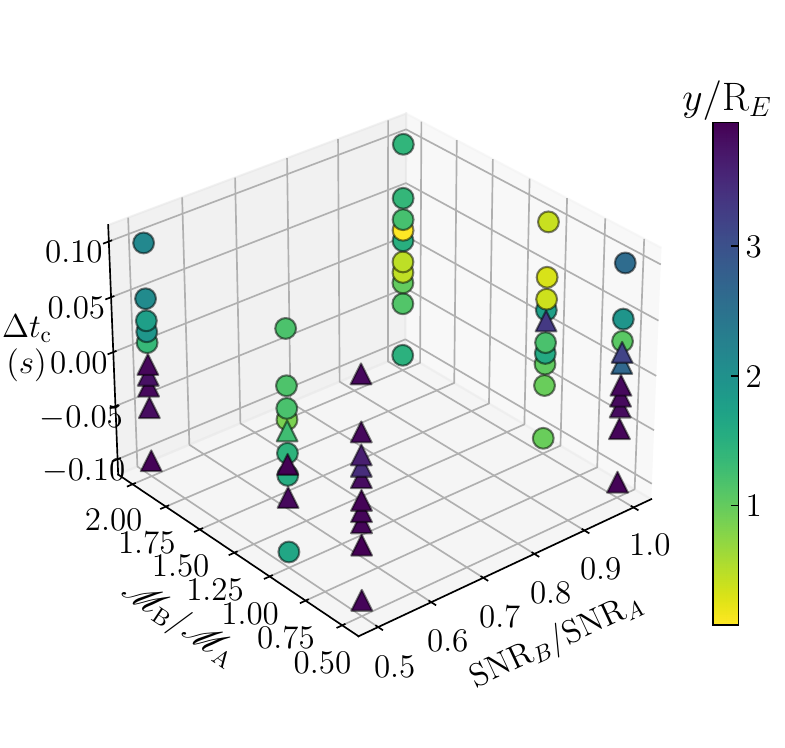}%
     \includegraphics[keepaspectratio, width=0.33\textwidth]{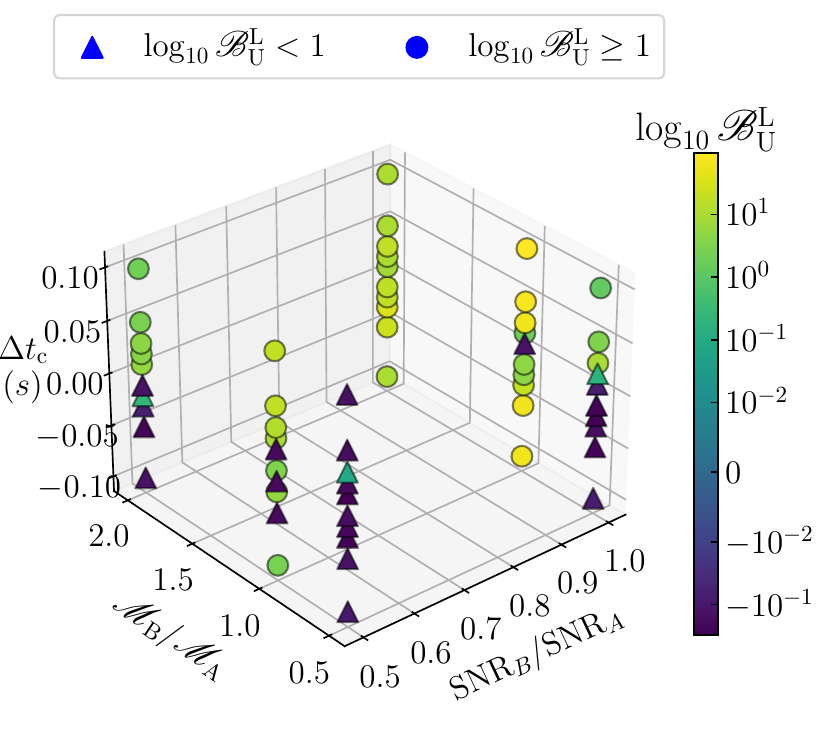}
     \caption{Inference of overlapping signals using microlensed templates for an isolated point-mass lens. \emph{Left:} Inferred logarithmic redshifted lens mass $\log_{10} M_{\rm L}^z/{\rm M}_{\odot}$. \emph{Middle:} Inferred dimensionless impact parameter $y/{\rm R_E}$. \emph{Right:} Bayes factor comparison between microlensed and unlensed hypotheses, with circular markers indicating $\log_{10}\mathscr{B}^{\rm L}_{\rm U} \geq 1$.}
     \label{fig:microlensedpeparams}
\end{figure*} 

\begin{figure}[htb!]
     \centering
     \includegraphics[keepaspectratio, width=0.485\textwidth]{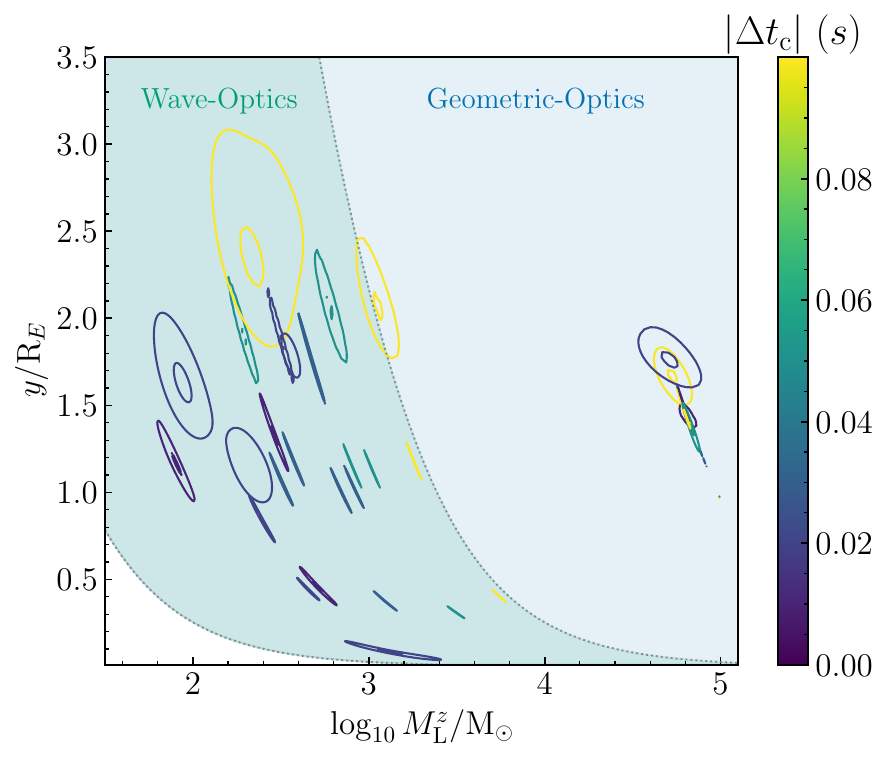}
     \caption{Joint 2D marginalized posterior distribution of the inferred logarithmic redshifted lens mass $M_{\rm L}^z$ and impact parameter $y$, assuming an isolated point-lens model, for the injected overlapping signals. The color scale encodes the coalescence time difference $|\Delta t_{\rm c}|$, and the level sets indicate the $50\%$ and $95\%$ confidence intervals. Only cases with substantial evidence in favor of microlensing ($\log_{10}\mathscr{B}^{\rm L}_{\rm U} > 0.5$) are shown for clarity. The shaded green (orange) region indicates where wave-optics (geometric-optics) effects are expected to dominate, determined by the inferred microlensing time delay between the images, $\tau(M_{\rm L}^z, y)$.}
     \label{fig:mlzylmicrolensedpe}
\end{figure} 

\begin{figure}[htb!]
     \centering
     \includegraphics[keepaspectratio, width=.485\textwidth]{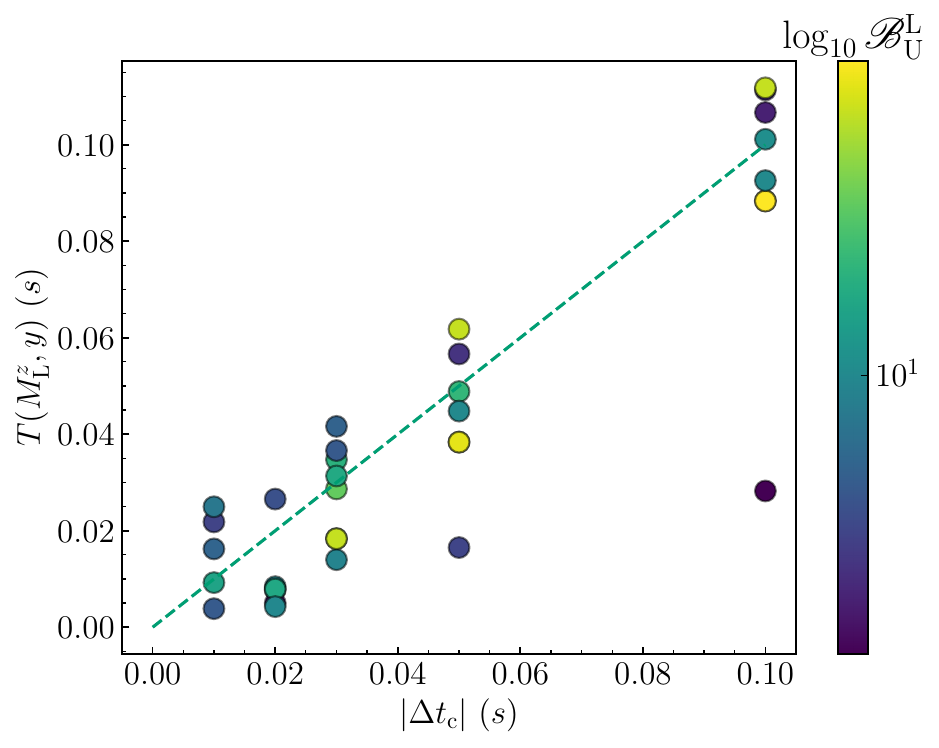}
     \caption{Comparison between the injected time difference $|\Delta t_{\rm c}|$ between the overlapping signals and the inferred lensing time delay $\tau(M_{\rm L}^z, y)$ between the microlensed images (Eq.~4.5 in Ref.~\cite{Nakamura:1999uwi}), for a lens system with redshifted lens mass $M_{\rm L}^z$ and impact parameter $y$ corresponding to the inferred PE values. Specifically, we compute $T(M_{\rm L}^z, y)=\min\left\{\tau(M_{\rm L}^z, y),\ 4 - \tau(M_{\rm L}^z, y)\right\}$ to account for the wrap-around effect arising from the finite $4~\rm{s}$ analysis duration used in PE, where the delayed image can appear before the primary one within the analysis window. The color scale represents $\log_{10}\mathscr{B}^{\rm L}_{\rm U}$, and only cases with substantial evidence for microlensing ($\log_{10}\mathscr{B}^{\rm L}_{\rm U} > 0.5$) are shown. The dashed green line indicates $|\Delta t_{\rm c}|=\tau(M_{\rm L}^z, y)$ for reference.}
     \label{fig:microlensedpetime}
\end{figure} 

\begin{figure*}[htb!]
     \centering
     \includegraphics[keepaspectratio, width=1\textwidth]{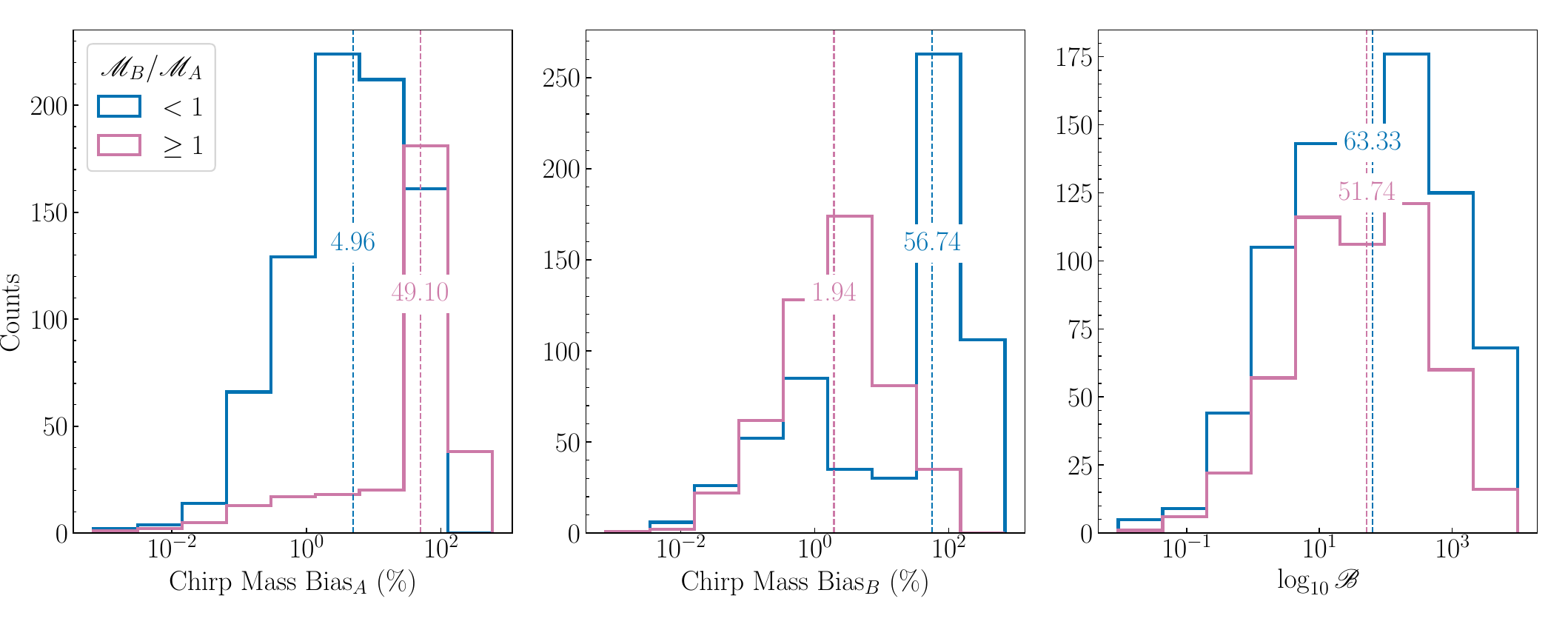}
     \caption{\emph{Left} and \emph{Middle:} Chirp mass biases, \emph{Right:} Bayes factors for a population of overlapping signals, obtained from fitting factor optimization with unlensed quasicircular templates. Bayes factor $\log_{10}\mathscr{B}$ quantifies the evidence in favor of an unlensed quasicircular signal over noise, estimated via Eq.~\eqref{eq:logbf_ff}. The histograms have been marginalized over $\mathrm{SNR}_{\rm B}/\mathrm{SNR}_{\rm A}$ and $\Delta t_{\rm c}$. The color represents differentiation for $\mathscr{M}_{\rm B}/\mathscr{M}_{\rm A}<1$ and $\mathscr{M}_{\rm B}/\mathscr{M}_{\rm A}\geq1$. Dashed lines denote the overall medians across the population.}
     \label{fig:singlesffpop}
\end{figure*}

\begin{figure}[htb!]
     \centering
     \includegraphics[keepaspectratio, width=0.485\textwidth]{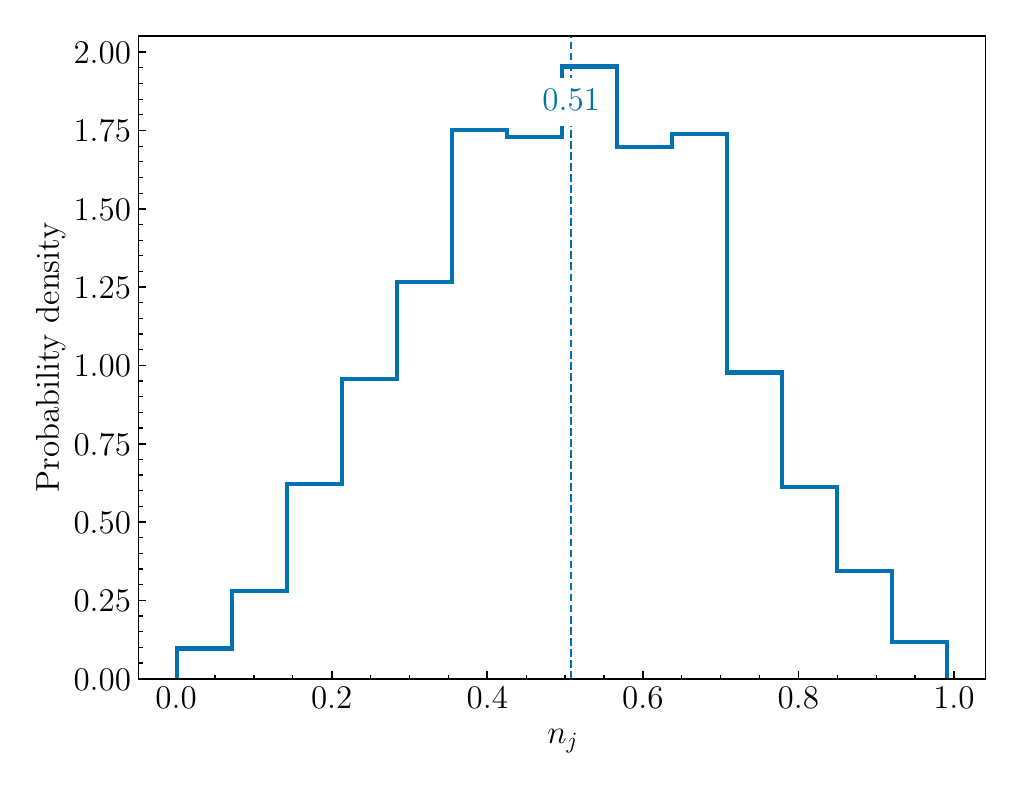}
     \caption{Population analysis of overlapping signals inferred with Type-II lensed templates. Results of the Morse phase densities when $n_j$ is varied to maximize the match. The histogram is marginalized over the ratios $\mathscr{M}_{\rm B}/\mathscr{M}_{\rm A}$, $\mathrm{SNR}_{\rm B}/\mathrm{SNR}_{\rm A}$ and $\Delta t_{\rm c}$ due to their similar distributions. Dashed lines denote the overall medians across the population.}
     \label{fig:slffpop}
\end{figure}
 
\begin{figure*}[htb!]
     \centering
     \includegraphics[keepaspectratio, width=1\textwidth]{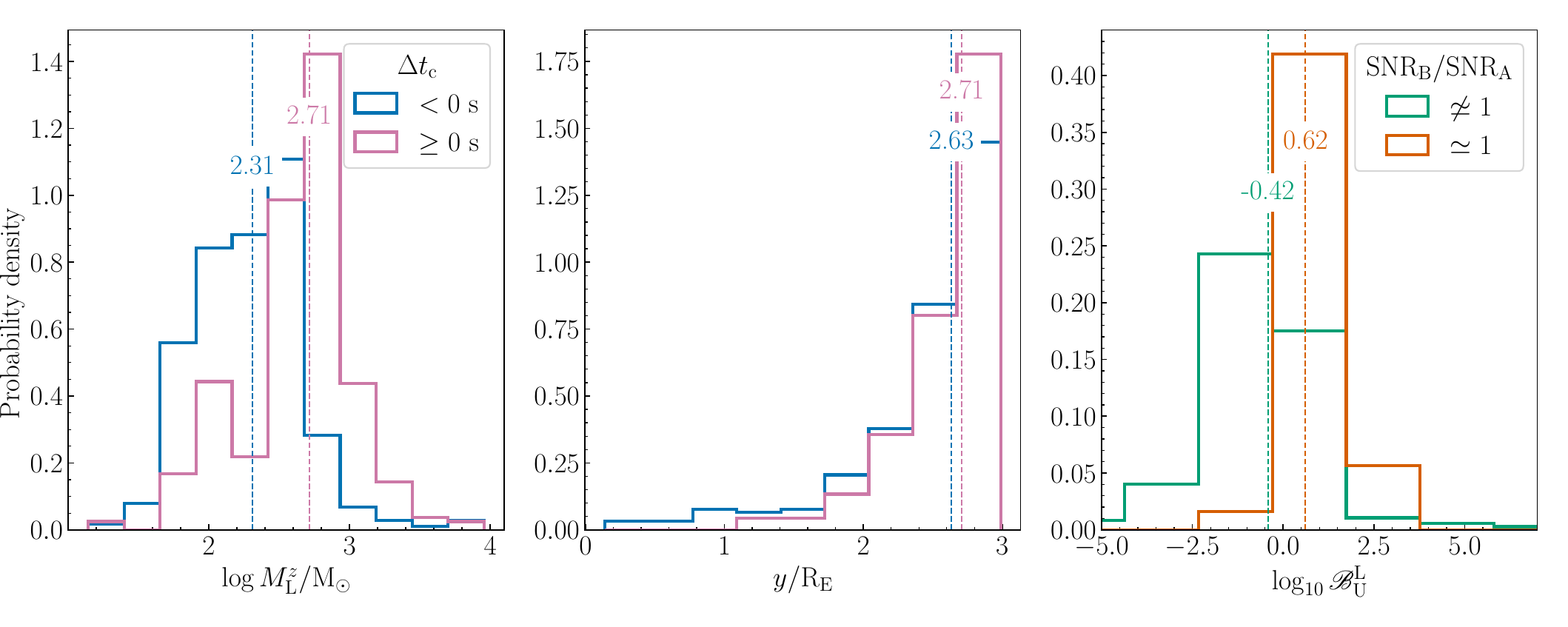}
     \caption{Fitting-factor results for overlapping signal populations using microlensed templates. \emph{Left:} Inferred redshifted lens mass $\log_{10} M_{\rm L}^z$ densities, \emph{Middle:} Impact parameter $y$ densities, \textit{Right:} Bayes factor support for microlensed model over unlensed, estimated using Eq.~\eqref{eq:logbf_ff}. The lens masses peak around $M_{\rm L}^z \sim 10^2 {\rm M}_\odot$ for negative coalescence time differences, and $M_{\rm L}^z \sim10^3 {\rm M}_\odot$ for positive $\Delta t_{\rm c}$, and we infer large impact parameters $y \sim 2.5$ for both cases. For the Bayes factor difference, the parameter space of interest was directed by the PE results in Sec.~\ref{subsubsec:mlpe}, against $\mathrm{SNR}_{\rm B}/\mathrm{SNR}_{\rm A}$, and marginalizing over $\mathscr{M}_{\rm B}/\mathscr{M}_{\rm A}$ and $\Delta t_{\rm c}$. Dashed lines denote the overall medians across the population.}
     \label{fig:microlensedffpop}
\end{figure*} 

\subsection{Parameter Estimation}
\label{subsec:resultspe}

We first analyze $60$ overlapping cases across the relative parameter space of overlapping signals. As mentioned previously in Sec.~\ref{sec:setups}, the SNR of ${\rm SINGLES_A}$ is fixed to $30$, with ${\rm SINGLES_A}$ always being the louder signal.

\paragraph*{Unlensed Quasicircular Model:} For the null hypothesis of an unlensed quasicircular GW signal, we observe that strong temporal overlap $|\Delta t_{\rm c}| < 0.03~{\rm s}$ frequently produces \emph{bimodal} posteriors in the inferred chirp mass, observed in $13$ of the $60$ cases. In such cases, the inferred chirp mass deviates from the chirp mass of either overlapping signal, and the inferred SNRs are also significantly lower. This suggests that strong overlap can induce degeneracies in the intrinsic mass parameters that the unlensed model cannot resolve. 

Fig.~\ref{fig:singlespe} summarizes the inferred chirp mass biases and the inferred SNRs for the unlensed runs. Since only $\mathrm{SNR}_{\rm A}$ is fixed in our injections, the total network SNR varies across the grid, both due to intrinsic changes in the injected signals and also the mismatches between the injections and the unlensed templates. This dependence is important because the Bayes factor scales steeply with SNR (Eq.~\ref{eq:snr_ff}); therefore, the results in Fig.~\ref{fig:singlespe} are directly relevant for interpreting the model comparison outcomes discussed later.

The inferred chirp mass for the louder signal ($\rm{SINGLES_A}$) is typically accurate to within $20\%$. In contrast, the weaker signal ($\rm{SINGLES_B}$) exhibits significantly larger systematic biases, up to $200\%$ in extreme cases. This asymmetry likely arises because the inferred geocentric time aligns more closely with that of the louder signal, yielding more accurate parameter estimates for $\rm{SINGLES_A}$. When the chirp mass ratios and SNRs of the two signals are comparable and the overlap is strong, the bias in the weaker signal’s chirp mass $(\mathscr{M}_{\rm B})$ becomes severe, and the effective SNR drops significantly. Even in cases with asymmetric chirp mass ratios but equal SNRs, we see gradations in inferences, indicating that the parameter degeneracies depend on both intrinsic and extrinsic factors.

\subsubsection{Type-II Lensed Image Model}
\label{subsubsec:slpe}

We next model the same set of overlapping signals with a strongly-lensed template corresponding to a Type-II image. Fig.~\ref{fig:slpeparams} compares results for two inference settings: a Type-II configuration i) with $n_j=0.5$ (left), and (ii) a free Morse index $n_j \in [0,1]$ (middle and right). In both cases, strong overlaps ($|\Delta t_{\rm c}| \lesssim 0.02~{\rm s}$) yield a mild but noticeable preference for Type-II lensing, indicated by positive values of lensed-to-unlensed Bayes factor, $\log_{10}\mathscr{B}^{\rm L}_{\rm U}$. For moderate to large temporal separations ($|\Delta t_{\rm c}| \gtrsim 0.05~{\rm s}$), or for strong SNR asymmetries, the preference disappears and the Bayes factors support the unlensed model. Consistent with previous studies~\cite{Jeffreys:1939xee, Kass:1995loi, Thrane:2018qnx, DelPozzo:2011pg}, we adopt $\log_{10}\mathscr{B}^{\rm L}_{\rm U} > 1$ (highlighted using circular markers in the relevant plots throughout this paper) as the threshold for strong support of the lensing hypothesis.

Fig.~\ref{fig:nslpe} shows that when $n_j$ is allowed to vary, both the maximum likelihood estimate (MLE) value and the median of the 1D marginalized posterior often cluster near 0.5 for overlapping injections\footnote{Since we use a uniform prior on $n_j$, the median of a weakly constrained posterior tends to lie near $0.5$. To mitigate this prior-induced bias, we also report MLE estimates.}, which indicates a Type-II lensed image. Wider time separations ($|\Delta t_{\rm c}| \gtrsim 0.02~{\rm s}$) cause the $n_j$ posterior to broaden or shift away from $0.5$, reflecting the fact that a single lensed image cannot fit distinct time–frequency features any better than the null hypothesis of a quasicircular unlensed signal. The fact that MLE values, which are independent of the prior volume effects, still peak near $0.5$ reinforces that this clustering is a genuine feature of the data in the high-overlap regime.

\subsubsection{Microlensed Model}
\label{subsubsec:mlpe}

We next investigate the ability of the microlensed model to fit overlapping signals. Fig.~\ref{fig:microlensedpeparams} shows the inferred microlensed parameters (left and middle) and the Bayes factors comparing the microlensed and unlensed hypotheses (right). The maximum support for the microlensing hypothesis occurs when $\mathscr{M}_{\rm B}/\mathscr{M}_{\rm A}=1$ and $\mathrm{SNR}_{\rm B}/\mathrm{SNR}_{\rm A}=1$, and this support increases as $|\Delta t_{\rm c}|$ moves away from zero, attaining a maximum at $|\Delta t_{\rm c}|=0.1~\mathrm{s}$. This trend is expected, since in the geometric-optics regime of the microlensed parameter space considered here, a microlensed signal is well approximated as the superposition of two images, with detector-frame chirp masses remaining invariant across lensed images. The rise in Bayes factors with increasing $|\Delta t_{\rm c}|$ can be understood if the microlens-induced effective time delay becomes comparable to the injected $\Delta t_{\rm c}$.

For $\mathrm{SNR}_{\rm B}/\mathrm{SNR}_{\rm A}=1$ and $\mathscr{M}_{\rm B}/\mathscr{M}_{\rm A}=2$, all values of $\Delta t_{\rm c}$ showed preference for microlensing. In contrast, for $\mathrm{SNR}_{\rm B}/\mathrm{SNR}_{\rm A}=\mathscr{M}_{\rm B}/\mathscr{M}_{\rm A}=0.5$ no case preferred microlensing. For the mixed cases $\mathrm{SNR}_{\rm B}/\mathrm{SNR}_{\rm A}=0.5$, $\mathscr{M}_{\rm B}/\mathscr{M}_{\rm A}=2$ and $\mathrm{SNR}_{\rm B}/\mathrm{SNR}_{\rm A}=1$, $\mathscr{M}_{\rm B}/\mathscr{M}_{\rm A}=0.5$ preference for microlensing was observed only when $\Delta t_{\rm c}>0$. The observed $\Delta t_{\rm c}$ asymmetry naturally follows from the fact that the microlensing model produces a positive effective time delay between images, so a better fit is expected when the sign and magnitude of the microlens delay align with the injected $\Delta t_{\rm c}$. $\Delta t_{\rm c} > 0$ naturally aligns with the lens model's causality, mimicking a standard demagnified secondary image. In contrast, cases with $\Delta t_{\rm c} < 0$ effectively present a scenario where the second signal is magnified relative to the first, which violates the causal ordering in lensing due to a point-mass lens.

Fig.~\ref{fig:mlzylmicrolensedpe} shows the inferred 2D marginalized joint posteriors for the redshifted lens mass and impact parameter versus coalescence time difference. Only cases with substantial evidence\footnote{We use Jeffreys' criterion~\citep{Jeffreys:1939xee} for Bayes factors. Since these injections are performed in zero noise, the Jeffreys' scale provides a conservative estimate, implying that a Bayes factor value of $0.5$ could correspond to a stronger preference for the alternative hypothesis in realistic noisy scenarios.} in favor of microlensing $(\log_{10} \mathscr{B}^{\rm L}_{\rm U} > 0.5)$ are shown for clarity. The inferred redshifted lens masses range from $\sim10^2$ to $10^5~{\rm M}_{\odot}$, with impact parameters varying between $\sim 0.1$ and $3~{\rm R_E}$. We note that as $|\Delta t_{\rm c}|$ decreases from $0.1~{\rm s}$ to $0.01~{\rm s}$, the inferred $\log_{10}M^{\rm z}_{\rm L}$ and $y$ values typically decrease, that is, from higher lens masses and impact parameters, indicative of the geometric optics regime, to lower values where wave optics effects dominate. This behavior is expected, since in the geometric optics approximation, a microlensed signal is well modeled as the superposition of two images separated by a time delay, $\tau(M^z_{\rm L}, y)$. 

We study this explicitly in Fig.~\ref{fig:microlensedpetime}, which shows the injected time difference $|\Delta t_{\rm c}|$ versus the inferred lensing time delay $\tau(M_{\rm L}^z, y)$ between the microlensed images (Eq.~4.5 in Ref.~\cite{Nakamura:1999uwi}), for a lens system with $M_{\rm L}^z$ and $y$ corresponding to the inferred maximum-a-posteriori (MAP) values. Specifically, we compare $|\Delta t_{\rm c}|$ with $T(M_{\rm L}^z, y)\equiv \min\{\tau(M_{\rm L}^z, y),\, 4 - \tau(M_{\rm L}^z, y)\}$ to account for the wrap-around effect arising from the finite $4~\rm{s}$ analysis duration used in the PE, where the delayed image can appear before the primary one within the analysis window. It is interesting to note that the inferred $T(M_{\rm L}^z, y)$ values typically lie close to the corresponding $|\Delta t_{\rm c}|$ values. This indicates that false evidence for microlensing signatures can arise because, to a reasonable approximation, the model produces two superimposed images whose time delay can closely match either the injected $|\Delta t_{\rm c}|$ or $(\delta - |\Delta t_{\rm c}|)$, where $\delta$ denotes the analysis duration ($=4~\rm{s}$) used in PE. The slight offsets between $|\Delta t_{\rm c}|$ and $T(M_{\rm L}^z, y)$ are expected and arise primarily due to three factors: (i) the intuition of overlapping images holds strictly in the geometrical-optics limit and may not fully capture wave-optics effects;(ii) systematic biases resulting from differences between the injection and recovery models; and (iii) inherent statistical uncertainties in the PE process.

Overall, the PE results show that both Type-II and microlensed models can provide modest improvements in fit over the unlensed model, though for different reasons in specific regions of the relative overlapping parameter space. This highlights the risk of misinterpreting overlapping events as gravitationally lensed.

\subsection{Fitting Factor Studies}
\label{subsec:ffpop}

Although Bayes-factor estimates derived from fitting factors in Eq.~\eqref{eq:logbf_ff} are most reliable in the high-SNR regime, and the best-fit template parameters may converge less robustly than the fitting factor itself, the approach provides a practical means to explore population-level trends. Preliminary results confirm that lower fitting factors correspond to reduced effective SNRs, which in turn degrade detection efficiency and bias parameter inference. 

We extended the fitting factor study to the population of overlapping signals drawn from realistic BBH distributions sampled from astropysically motivated priors. Fig.~\ref{fig:singlesffpop} summarizes results for unlensed quasicircular templates. We have an asymmetry in the relative chirp mass biases inferred, with respect to $\rm{SINGLES_A}$ and $\rm{SINGLES_B}$, likely due to the geocentric time of the higher SNR signal ($\rm{SINGLES_A}$) aiding the inference. We have marginalized the representations over $\mathrm{SNR}_{\rm B}/\mathrm{SNR}_{\rm A}$ and $\Delta t_{\rm c}$, since the distributions are similar, with the significant parameter of interest being the chirp mass ratio $\mathscr{M}_{\rm B}/\mathscr{M}_{\rm A}$. Our studies align with previous works for the chirp mass biases, and we have higher biases for $\rm{SINGLES_B}$ at lower $\mathscr{M}_{\rm B}/\mathscr{M}_{\rm A}$, and vice-versa. The Bayes factor $\log_{10}\mathscr{B}$ provides an estimate from Eq.~\eqref{eq:logbf_ff}, which seems to have similar distributions for smaller and larger chirp mass ratios.

\subsubsection{Type-II Lensed Template}
\label{subsubsec:slffpop}
 
Fig.~\ref{fig:slffpop} presents fitting factor results for the Type-II lensing hypothesis. When the Morse index is allowed to vary, the inferred index typically clusters around $n_j\simeq0.5$ in the most overlapping scenarios. This is consistent with the PE results, suggesting a signature of Type-II Lensing. We don't observe any specific trend in the relative parameter space of the overlapping signals; hence, the histograms are marginalized over the intrinsic parameters of interest. Although not shown here, the Bayes factors estimated from the FF show a significant peak around $0$, with mild support when the Morse index varies. 

\subsubsection{Microlensed Template}
\label{subsubsec:mlffpop}

Population results for microlensed templates are shown in Fig.~\ref{fig:microlensedffpop}. The inference across the population has moderate lens masses, peaking in the range of $10^2$--$10^3~{\rm M}_{\odot}$, and large impact parameters, $y>2~{\rm R_E}$. We observe two different peaks in the lens mass, with moderate lens masses ($\sim10^2~{\rm M}_{\odot}$) being inferred by negative $\Delta t_{\rm c}$, and larger values ($\sim10^3~{\rm M}_{\odot}$) seen in $\Delta t_{\rm c}>0$, similar to the trends observed for the PE inferences in Sec.~\ref{subsubsec:mlpe}, specifically in Fig.~\ref{fig:microlensedpeparams}. The Bayes factors are predominantly clustered around $0$, with a subtle preference for positive Bayes factors for $\mathrm{SNR}_{\rm B}/\mathrm{SNR}_{\rm A} \sim 1$, for a symmetric SNR of both signals. This suggests that lensing is likely degenerate with overlapping signals, when two signals of similar loudness interfere, which aligns with the inferences from the PE results in Sec.~\ref{subsubsec:mlpe}. The FF analysis thus confirms that the degeneracy between microlensing and overlapping signals is confined to a specific and narrow region of the parameter space, characterized by comparable SNRs. In all other regimes, the unlensed hypothesis remains the preferred explanation, with slight negative Bayes factor recoveries, especially for disparate loudness.

\section{Conclusions}
\label{sec:conclusions}

With increasing sensitivity of the current and future GW detectors, the likelihood of observing two unrelated overlapping GW signals or signals that are gravitationally lensed will rise. In this work, we have investigated the potential of such overlapping signals to mimic the modulations due to gravitational lensing, focusing on two scenarios: \emph{Type-II} strong lensing and \emph{microlensing} in the wave-optics regime. 

We generated a simulated zero-noise population, each comprising two overlapping GW signals with varied chirp mass ratios $\mathscr{M}_{\rm B}/\mathscr{M}_{\rm A}$, SNR ratios $\mathrm{SNR}_{\rm B}/\mathrm{SNR}_{\rm A}$, and coalescence time differences $\Delta t_{\rm c}=t_{\rm c}^B-t_{\rm c}^A$. To explore degeneracies between overlapping binaries and lensed signals, we employ two complementary approaches: (i) full Bayesian PE using \texttt{dynesty} nested sampler as implemented in \texttt{BILBY}, and (ii) fitting factor optimization to identify the best-matching template. Both methods were used for Bayesian model selection, comparing the lensed and unlensed hypotheses via the Bayes factors defined in Eqs.~\eqref{eq:logbf} and \eqref{eq:logbf_ff}.

A targeted PE analysis was first performed on $60$ injections spanning $\mathscr{M}_{\rm B}/\mathscr{M}_{\rm A} \in \{0.5, 1, 2\}$, $\mathrm{SNR}_{\rm B}/\mathrm{SNR}_{\rm A} \in \{0.5, 1\}$, and $\Delta t_{\rm c} \in [-0.1, 0.1]~\mathrm{s}$, using lensed templates. We then extended the analysis to population studies of $\mathcal{O}(5000)$ signals, using FF optimization for computational efficiency. \

The main findings are:
\begin{itemize}
    \item{\emph{Type-II Lensed (PE)}:} 
    A preference for Type-II lensing ($\log_{10} \mathscr{B}^{\rm L}_{\rm U} > 1$) arises only in a small region of parameter space, $|\Delta t_{\rm c}|\leq 0.03~\mathrm{s}$. The median inferred Morse index is around $n_j \simeq 0.5$, consistent with the saddle-point image of Type-II lensing.
    
    \item{\emph{Type-II Lensed (FF)}:}
    The support for the Type-II lensed hypothesis is confined to a small region of the population parameter space of interplay of $\mathscr{M}_{\rm B}/\mathscr{M}_{\rm A}$ and the $|\Delta t_{\rm c}|$, with the inferred Morse index clustered around $n_j\simeq0.5$, and does not provide any clear trend with any of the relative parameters. The FF results are consistent with the trends observed in the PE analyses.
    
    \item{\emph{Microlensing (PE)}:}
    The false evidence for microlensing signatures generally arises because the model can, to a reasonable approximation, produce two superimposed images with time delay closely matching either $|\Delta t_{\rm c}|$ or $(\delta - |\Delta t_{\rm c}|)$, where $\delta$ is the analysis duration ($4~\rm{s}$) used in PE. We infer redshifted lens masses in the range $M_{\rm L}^z\sim10^2$--$10^{5}~{\rm M}_{\odot}$ and impact parameters $y\sim0.1$--$3~{\rm R_E}$. We note that as $|\Delta t_{\rm c}|$ decreases from $0.1~{\rm s}$ to $0.01~{\rm s}$, the inferred $\log_{10}M^{\rm z}_{\rm L}$ and $y$ values typically decrease, from the geometric optics regime, to where wave optics effects dominate. The maximum support occurs for $\mathscr{M}_{\rm B}/\mathscr{M}_{\rm A}=\mathrm{SNR}_{\rm B}/\mathrm{SNR}_{\rm A}=1$, and this support increases as $|\Delta t_{\rm c}|$ increases. For $\mathrm{SNR}_{\rm B}/\mathrm{SNR}_{\rm A}=1$ and $\mathscr{M}_{\rm B}/\mathscr{M}_{\rm A}=2$, all $\Delta t_{\rm c}$ values preferred microlensing.

    \item{\emph{Microlensing (FF)}:}
    FF optimization shows moderate microlensing support when $\mathrm{SNR}_{\rm B}/\mathrm{SNR}_{\rm A} \sim 1$, consistent with PE trends. Outside this region, we have negative Bayes factors, and the unlensed model is preferred.
\end{itemize}

These results highlight that overlaps between two black hole binaries can produce biased parameter inference and false support for lensing hypotheses, particularly mimicking microlensing. False support for Type-II lensing is minimal, occurring only for very strongly overlapping cases with $|\Delta t_{\rm c}|\lesssim 0.03~{\rm s}$. In contrast, false evidence for microlensing generally arises when the model produces two images whose time delay matches either the injected $|\Delta t_{\rm c}|$ or $(\delta - |\Delta t_{\rm c}|)$. The inferred lens masses and impact parameters depend on the temporal overlap, with maximal support for equal-mass, equal-SNR binaries. These degeneracies can reduce the effective SNR, degrade detection efficiency, and increase the risk of misclassification.

In the future, the growing sensitivity of GW detectors will increase the probability of overlaps, especially for long-duration signals. Accounting for the possibility of overlapping events will therefore be essential before attributing waveform modulations to lensing. For widely separated signals, joint parameter estimation of the two overlapping signals would be efficient in separating the intrinsic parameter spaces and overlap. 
Randomly overlapping binaries are distinct astrophysical events that will generally possess disjoint sky location posteriors. In contrast, true gravitationally lensed images must originate from the same sky location to within the resolution of the detector network. As such, consistency in sky localization will thus serve as a crucial discriminator in unraveling the degeneracies identified in this work. In our study, however, we consider strong overlaps with $|\Delta t_{\rm c}|\lesssim0.1~{\rm s}$, which may not be detected as separate signals by search pipelines and hence, the sky localization information cannot be used as an additional constraint. Such tightly overlapping signals could be interpreted as a single signal, and we might mistakenly attribute overlapping features to lensing, especially when the second signal is below the detection threshold. Moreover, overlap-induced waveform interference can mimic other physical effects: in particular, the modulation can be qualitatively similar to the signature of orbital eccentricity. We will address this degeneracy between overlapping and eccentric signals in our future work, aiming to disentangle these scenarios and ensure robust astrophysical interpretation of forthcoming GW detections.

\section*{Acknowledgements}
The authors are thankful to H. Narola for reviewing the manuscript during the LSC Publications and Presentations procedure and providing useful comments. 
We thank P. Relton and J. Janquart for valuable discussions and feedback. 
The authors are also grateful for computational resources provided by the LIGO Laboratory and supported by National Science Foundation Grants PHY-0757058 and PHY-0823459, and acknowledge the use of the IUCAA LDG cluster Sarathi for computational work.
N. Rao further acknowledges support from KVPY for funding. 
This material is based upon work supported by NSF's LIGO Laboratory, which is a major facility fully funded by the National Science Foundation.
The research of A. Mishra is supported by the Department of Atomic Energy, Government of India, under Project No. RTI4001.

The work utilizes the following software packages:
\texttt{Cython}~\cite{cython},
\texttt{NumPy}~\cite{Harris:2020xlr}, 
\texttt{SciPy}~\cite{Virtanen:2019joe}, 
\texttt{PyCBC}~\cite{pycbc},
\texttt{LALSuite}~\cite{lalsuite,swiglal},
\texttt{dynesty}~\cite{Speagle:2019ivv}, 
\texttt{BILBY}~\cite{Ashton:2018jfp,Smith:2019ucc,Romero-Shaw:2020owr},
\texttt{corner}~\cite{corner},
\texttt{gwmat}~\cite{Mishra:2023ddt},
\texttt{gwpopulation}~\cite{Talbot:2019okv},
\texttt{Matplotlib}~\cite{Hunter:2007ouj}, and
\texttt{Jupyter notebook}~\cite{jupyter}.

\section*{Data Availability}
The data and analysis codes used in this work are publicly made available at the repository on Github~\cite{git_overlap}.

\bibliography{bibliography}% Produces the bibliography via BibTeX.

\end{document}